\documentclass[manuscript,screen]{acmart}

\AtBeginDocument{%
  \providecommand\BibTeX{{%
    \normalfont B\kern-0.5em{\scshape i\kern-0.25em b}\kern-0.8em\TeX}}}

\usepackage{graphicx}
\usepackage[caption=false]{subfig}
\usepackage{hyperref}
\usepackage{multirow}
\usepackage{newunicodechar}
\newunicodechar{≤}{\ensuremath{\leq}}

\begin{document}

\title{An Analysis of Various Design Pathways Towards Multi-Terabit Photonic  On-Interposer Interconnects}


\author{Venkata Sai Praneeth Karempudi}
\affiliation{%
  \institution{University of Kentucky}
  \city{Lexington}
  \state{Kentucky}
  \country{USA}}
\email{kvspraneeth@uky.edu}

\author{Janibul Bashir}
\affiliation{%
  \institution{National Institute of Technology, Srinagar}
  \city{Srinagar}
  \country{India}}

\author{Ishan G Thakkar}
\affiliation{%
 \institution{University of Kentucky}
 \city{Lexington}
 \state{Kentucky}
 \country{USA}}

\renewcommand{\shortauthors}{Karempudi, et al.}

\begin{abstract}
In the wake of dwindling Moore's Law, to address the rapidly increasing complexity and cost of fabricating large-scale, monolithic systems-on-chip (SoCs), the industry has adopted dis-aggregation as a solution, wherein a large monolithic SoC is partitioned into multiple smaller chiplets that are then assembled into a large system-in-package (SiP) using advanced packaging substrates such as silicon interposer. For such interposer-based SiPs, there is a push to realize on-interposer inter-chiplet communication bandwidth of multi-Tb/s and end-to-end communication latency of no more than 10 ns. This push comes as the natural progression from some recent prior works on SiP design, and is driven by the proliferating bandwidth demand of modern data-intensive workloads. To meet this bandwidth and latency goal, prior works have focused on a potential solution of using the silicon photonic interposer (SiPhI) for integrating and interconnecting a large number of chiplets into an SiP. Despite the early promise, the existing designs of on-SiPhI interconnects still have to evolve by leaps and bounds to meet the goal of multi-Tb/s bandwidth. However, the possible design pathways, upon which such an evolution can be achieved, have not been explored in any prior works yet. In this paper, we have identified several design pathways that can help evolve on-SiPhI interconnects to achieve multi-Tb/s aggregate bandwidth. We perform an extensive link-level and system-level analysis in which we explore these design pathways in isolation and in different combinations of each other. From our link-level analysis, we have observed that the design pathways that simultaneously enhance the spectral range and optical power budget available for wavelength multiplexing can render aggregate bandwidth of up to 4 Tb/s per on-SiPhI link. We also show that such high-bandwidth on-SiPhI links can substantially improve the performance and energy-efficiency of the state-of-the-art CPU and GPU chiplets based SiPs.
\end{abstract}



\maketitle

\section{Introduction}
With the recent deluge of data-centric computing applications, such as deep learning, and graph analytics, the world’s appetite for analyzing massive amounts of structured and unstructured data has grown dramatically. For instance, since 2012, the amount of compute used in the largest AI training jobs has been increasing exponentially with a 3.4-month doubling time \cite{AIandcompute}, which is 50$\times$ faster than the pace of Moore's Law. Fulfilling this appetite demands for increasingly high computational capacity (in terms of compute and memory bandwidths) and energy efficiency. However, consistently meeting  this sustaining demand by using the currently utilized large-scale computing systems, which typically employ a combination of large monolithic manycore chips and homogeneous multi-chip board designs (e.g., \cite{chen2014}\cite{chung2017}\cite{fowers2018}\cite{gwennap2018}\cite{jouppi2017}\cite{venkataramani2017}), is becoming increasingly very difficult due to three fundamental reasons. First, this demand is quickly outpacing the progress realized by dwindling Moore’s law, due to the fundamental physical limitations slowing the rate and increasing the complexity and cost of transition from one technology node to the next \cite{semiconductor}. Second, the attempts to scale the size of large monolithic chips gives rise to extravagant manufacturing cost due to the limited reticle size and poor yield of stitching multiple reticles together \cite{naffziger2021pioneering}. Third, scaling the multi-chip board designs can push the package to die ratio in such designs to be greater than 10:1 \cite{sodani2016}, which in turn can dramatically increase the area overhead of computing systems that employ such multi-chip board designs.

To overcome these challenges, the industry has focused on system dis-aggregation as a solution, wherein a large monolithic system-on-chip is partitioned into multiple smaller, modular chiplets of heterogeneous types. These chiplets are then assembled into a large system-on-package using organic substrate (e.g., \cite{iyer2016}\cite{arunkumar2017}), silicon interposer (e.g., \cite{stow2017}\cite{kannan2015}\cite{jerger2014noc}\cite{hu2016system}\cite{lai2016}), or silicon wafer (e.g., \cite{pal2019}\cite{pal2021}\cite{pal2018}\cite{bajwa2017}\cite{jangam2017}) as the substrate for chiplet assembly and packaging. The size of the silicon interposer based chiplet assemblies is typically limited to <1,000 mm$^2$ due to the limited reticle size \cite{naffziger2021pioneering}. Nevertheless, the silicon interposer based chiplet assemblies have several advantages over the organic substrate and silicon wafer based assemblies. Unlike the organic substrate based assemblies, the silicon interposer based assemblies have lower package-to-die ratio \cite{sodani2016}, which decreases their system area overheads. Along the same lines, unlike the silicon wafer based assemblies, the silicon interposer based assemblies are relatively less susceptible to challenges related to power delivery and thermal stability. Moreover, the silicon interposer based assemblies can also provide opportunities to have the active front-end-of-line logic components directly integrated onto the silicon interposer \cite{stow2017}, which provides more opportunities to increase the bandwidth density and efficiency of inter-chiplet interconnects by making it possible to implement advanced network topologies and routing logic directly on the interposer. On the other hand, the waferscale chiplet assemblies are very large in size compared to the interposer based assemblies. But unlike the silicon interposer, to achieve >90\% yield, the silicon wafer substrate, upon which various chiplets are assembled, has to remain passive. Because of these advantages, the silicon interposer based chiplet assemblies are rapidly materializing in both the industry and academia. 

As such, silicon interposer based chiplet assemblies require efficient implementation of inter-chiplet communication with low end-to-end latency, high bandwidth density, and high scalability, all achieved within a strict power budget. In general, the silicon interposer substrate can be active or passive \cite{stow2017}, and its use for assembling chiplets can be based on through silicon vias (TSVs) (e.g., as in TSMC's CoWoS technology family \cite{chuang2013}) or completely free of TSVs (e.g., as in Intel's EMIB technology family \cite{mahajan2016}). Regardless of the type of the utilized interposer, the interposer based chiplet systems can support inter-chiplet interconnects with tangibly very high (potentially multi-Tb/s) bandwidth densities \cite{Intel}. But such extreme-scale interconnect bandwidths are supported only for the inter-chiplet distance of less than 200-300 $\mu$m. In addition, prior work \cite{bharadwaj2020} has shown that as the number of chiplets on the interposer increases, the average latency of the inter-chiplet interconnects in the state-of-the-art interposer assemblies scales very poorly, regardless of the utilized interconnects topology. This is mainly because the data rates and latency of on-interposer electrical wires scale poorly due to their high impedance dependence.  To overcome these shortcomings, prior works have proposed active silicon-photonic interposer (SiPhI) based chiplet systems (e.g., \cite{bashir2017}\cite{thonnart2020}). These systems consider the bandwidth density of the inter-chiplet on-SiPhI interconnects to be $\sim$1 Tb/s/mm$^2$, because there is a push for the next generation interconnects to have >1 Tb/s/mm$^2$ bandwidth density \cite{DARPA}. In fact, SiPh based chiplet assemblies from prior work have shown multi-Tb/s/mm$^2$ bandwidth densities for optical fiber based off-package I/Os \cite{daudlin2021}\cite{rakowski2018}. As the natural progression from these excellent outcomes from prior works and driven by the increasing bandwidth needs of emerging workloads, there is impetus to  achieve multi-Tb/s bandwidth across the SiPh interposer with an end-to-end latency of no more than $\sim$10ns. However, to meet this goal, there are some daunting challenges to overcome. The major challenge is that as a SiPhI system scales to reach the reticle limit, the length of the end-to-end on-SiPhI links tends to become greater than 10 cm. For such long links, optical signal losses can become notably high, which in turn can make it very difficult to achieve even multi-Tb/s interconnect bandwidth, let alone achieving multi-Tb/s/mm$^2$ bandwidth density. Unfortunately, this challenge has not been addressed by any prior works so far.

To address this challenge, in this paper, \textit{for the first time}, we identify the key pathways for the design of multi-Tb/s on-SiPhI links, by taking clues from the existing literature on the design and optimization of SiPh interconnects, both in the on-chip and off-chip design domains. Our identified design pathways include: (1) increasing the available optical power budget per on-SiPhI link by minimizing the insertion losses and power penalties in the link, (2) increasing the spectral bandwidth available per on-SiPhI link (normally referred to as free-spectral range (FSR)) for higher degree of wavelength multiplexing, and (3) increasing the available optical power budget per on-SiPhI link by increasing the maximum allowable optical power (MAOP) limits of the link. We explore these SiPhI link-level design pathways in isolation and in various combinations of one another, to investigate which of these design pathways can help achieve multi-Tb/s on-SiPhI links. Based on our link-level analysis, We also enable the following two chiplet-based systems with our designed on-SiPhI multi-Tb/s links and provide their system-level performance analysis: \textit{(i)} a CPU based manycore multi-chiplet architecture named NUPLet \cite{bashir2017}, and \textit{(ii)} a GPU based deep learning training system from \cite{khani2021} that employs a total of 512 multi-chiplet GPU modules. 

The key contributions of this paper are summarized below: 

\begin{itemize}
\item We consider three state-of-the-art SiPh fabrication platforms from prior works \cite{stojanovic2018} and \cite{atabaki2018}, and then derive different variants of on-SiPhI links based on different combinations of these considered platforms and our identified design pathways mentioned earlier; 
\item We perform link-level analysis for all the derived on-SiPhI link variants, from which we calculate the achievable aggregate bandwidth and energy-per-bit (EPB) values for each on-SiPhI link variant for link lengths of up to 10 cm;
\item We identify all viable on-SiPhI link variants that can support multi-Tb/s aggregate bandwidth; 
\item We use our identified viable link variants to enable and evaluate different variants of two SiPhI based multi-chiplet systems from prior work: (1) a CPU system from \cite{bashir2017}, and (2) a GPU based deep learning training system based on \cite{khani2021};
\item We perform benchmark-driven analysis of our considered CPU based system variants to evaluate their performance (in terms of execution time), energy and energy-delay product, for PARSEC benchmark applications. Similarly, we also analyze our considered GPU based system variants to evaluate the training time-to-accuracy for deep learning applications. 
\end{itemize}

\section{Preliminaries}
\subsection{On-Silicon Photonic Interposer (On-SiPhI) Inter-Chiplet Links}
\begin{figure}
    \centering
    \includegraphics[scale = 0.11]{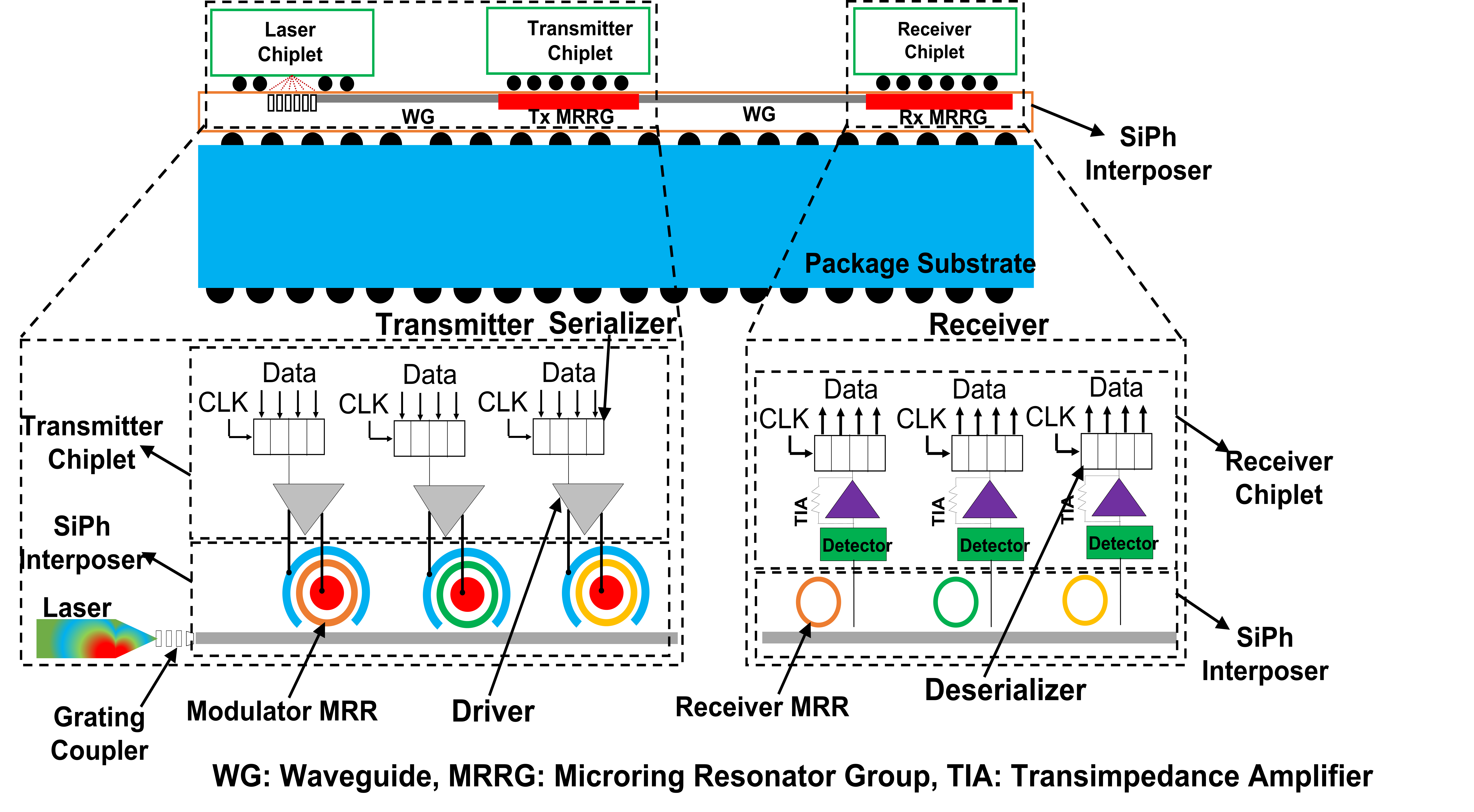}
    \caption{Inter-Chiplet Silicon Photonic MRRbased DWDM Link.}
    \label{Fig:1}
\end{figure}

Prior work \cite{abrams2020} provides a survey of design methods for multi-chiplet packages that integrate silicon photonics and electronics together. The use of a SiPhI for such integration is one of the approaches advocated in this work. Based on this approach, Fig. \ref{Fig:1} illustrates our envisioned schematic of an on-SiPhI link. The basic component of an on-SiPhI link is a silicon waveguide (shown in gray in Fig. \ref{Fig:1}) that is implemented on the SiPhI. The other SiPh components of the link that are implemented on the SiPhI include: a grating coupler; a transmitter microring resonator group (Tx MRRG); and a receiver microring resonator group (Rx MRRG). In addition, the on-SiPhI link also has other electronic and electro-optic components that are implemented on chiplets. These components are: a laser chiplet that has a comb laser source implemented on it \cite{stern2018}\cite{gaeta2019}; a transmitter chiplet that has Tx MRRG peripheral circuits such as modulator drivers and  serializers; and a receiver chiplet that has Rx MRRG peripheral circuits such as transimpedance amplifiers, and deserializers. The comb laser source on the laser chiplet emits a comb of optical wavelengths that are coupled into the on-SiPhI silicon waveguide via the grating coupler using the dense wavelength division multiplexing (DWDM) technique. These wavelengths work as different data-carrying channels. When these wavelength channels reach the Tx MRRG, the individual MRR modulators of the Tx MRRG modulate input data signals onto these wavelength channels. These modulated wavelength channels are transmitted to the Rx MRRG at the receiver side through the on-SiPhI silicon waveguide. The Rx MRRG consists of an array filter MRRs whose resonances are tuned to the incoming wavelength channels. These MRRs drop the incoming modulated channels onto their respective photodetectors to recover the input data signals. If the number of multiplexed wavelength channels into the on-SiPhI waveguide is N$_\lambda$, and if each wavelength channel operates at bitrate of \textit{BR} Gb/s, then the on-SiPhI waveguide can support N$_\lambda$ $\times$ \textit{BR} Gb/s bandwidth. Hence, to achieve >1 Tb/s bandwidth, the on-SiPhI waveguide must support sufficiently high values of N$_\lambda$ and \textit{BR}. Factors that impact the achievable values of N$_\lambda$ and \textit{BR} per on-SiPhI waveguide are discussed next.

\begin{figure}
    \centering
    \includegraphics[scale = 0.13]{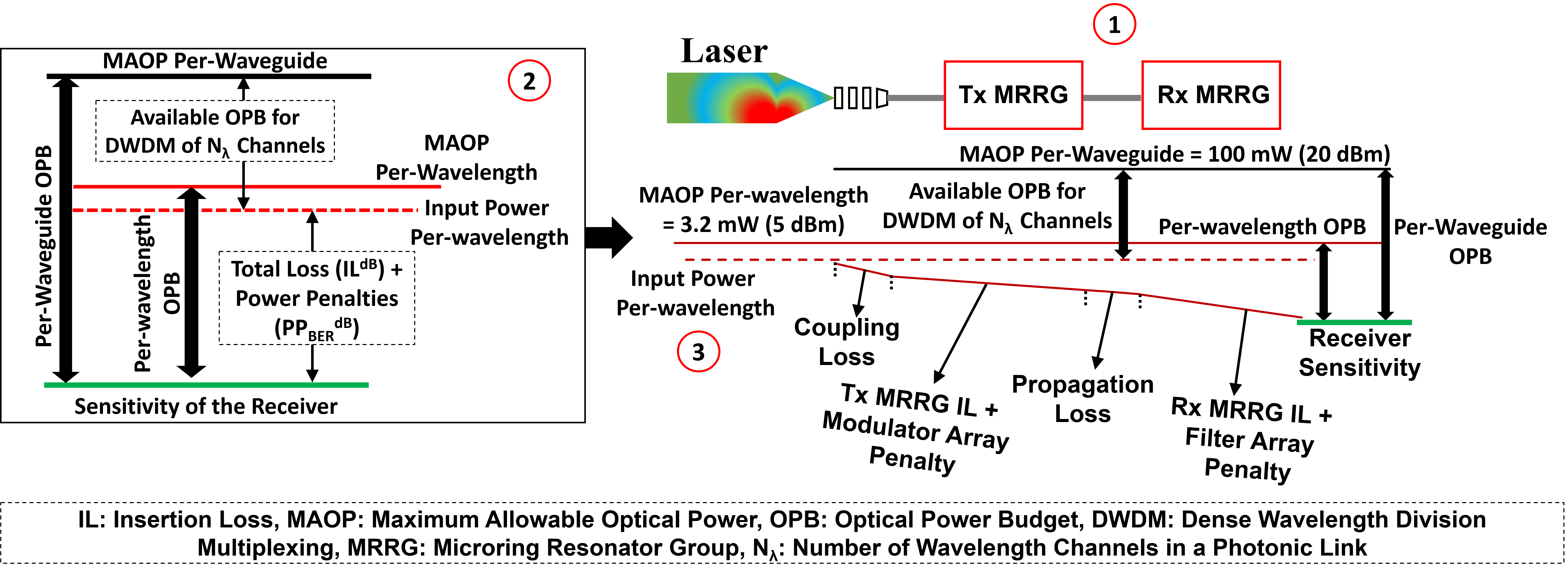}
    \caption{(1) Schematic of an on-SiPhI inter-chiplet link, (2) evolution of optical power budget and (3) summary of optical power budget}
    \label{Fig:2}
\end{figure}

\subsection{Performance of On-SiPhI Links}
It is well established in prior works that the performance (i.e., achievable N$_\lambda$ and \textit{BR}) of a SiPh link, whether an optical fiber based off-chip link (e.g., \cite{bahadori2016design}\cite{bahadori2017}\cite{bahadori2018}) or a silicon waveguide based on-chip link (e.g., \cite{sai2021}\cite{vatsavai2020}\cite{thakkar2017}\cite{bashir2019}\cite{pasricha2020}), depends on the strict optical power budget (OPB) of the link. This holds true for the on-SiPhI waveguide based links too. In this section, we refer to Fig. \ref{Fig:2} to illustrate how the OPB of an on-SiPhI link impacts its performance, i.e., its achievable N$_\lambda$ and \textit{BR}. 
As illustrated in (3) of Fig. \ref{Fig:2}, the OPB of a link determines the apex of the allowable optical losses and power penalties in the link. The OPB of a link has two mutually related components (see (2) in Fig. \ref{Fig:2}): (i) per-wavelength OPB, and (ii) per-waveguide OPB. Per-wavelength OPB determines the amount of allowable losses and power penalties for a single wavelength channel in the link, and can be defined as the difference between the per-wavelength maximum allowable optical power (MAOP) and the sensitivity of the receiver (Eq. (1)). Similarly, per-waveguide OPB determines the amount of optical losses and power penalties allowed for all the wavelength channels in the link, and it is provided as the difference between the maximum allowable optical power (MAOP) per waveguide and the sensitivity of the receiver (Eq. (2)). As illustrated in (2) of Fig. \ref{Fig:2}, the per-wavelength MAOP is restricted to 3.2 mW (5 dBm) (i.e., no more than 5 dBm optical power per wavelength is allowed). This limit has been decided upon theoretically \cite{bahadori2017}\cite{bahadori2018}\cite{wang2018} as well as empirically \cite{Qi2012}\cite{luo2012} to avoid the MRR modulators of the Tx MRRG from being inoperative due to the adverse impacts of optical non-linear effects such as multi-stability, self-heating, self-pulsation \cite{de2019}\cite{borghi2021}. On the other hand, the per-waveguide MAOP is restricted to 100 mW (20 dBm), to avoid dramatically high optical propagation losses in on-SiPhI waveguides caused due to the increased two-photon absorption (TPA) and free-carrier absorption (FCA) \cite{hendry2014}\cite{bahadori2017}\cite{thakkar2016}\cite{sai2021}.

\begin{equation}
        OPB\ per\ Wavelength\ (dB) = MAOP\ per\ Wavelength - Receiver\ Sensitivity
\end{equation}

\begin{equation}
        OPB\ per\ Waveguide\ (dB) = MAOP\ per\ Waveguide - Receiver\ Sensitivity
\end{equation}

Looking at the evolution of OPB provided in (2) of Fig. \ref{Fig:2}, a wavelength channel generated from a laser source experiences insertion loss and other power penalties as it propagates through the on-SiPhI waveguide of the link. The total insertion loss experienced by a wavelength channel includes: (i) the total coupling loss of the grating coupler; (ii) the waveguide propagation loss, which is the sum of the scattering loss (due to the sidewall roughness of the on-SiPh waveguide) and absorption loss (due to the material and free-carrier absorption mechanisms in the on-SiPh waveguide); and (iii) insertion loss of Tx+Rx MRRGs. On the other hand, the power penalties experienced by a wavelength channel across the link include the modulator array penalty (i.e., the power penalty incurred due to the array of modulator MRRs of the Tx MRRG) and detector array penalty (i.e., the power penalty incurred due to the array of filter MRRs of the Rx MRRG), as shown in Fig. \ref{Fig:2}. The modulator array penalty consists of modulator inter-channel crosstalk \cite{bahadori2016}. Similarly, the filter array penalty consists of the total power penalty manifesting at the photodetectors due to the inter-channel crosstalk at the MRR filters and truncation of the modulated signal spectra \cite{bahadori2016}. All of these optical insertion losses (IL$^{dB}$ in Eq. (3)) and power penalties (PP$_{BER}^{dB}$ in Eq. (3)) in the link as a whole (P$_{loss}^{dB}$ in Eq. (3)) should amount to be less than the per-wavelength OPB (Eq. (4)), for the link to be viable. This P$_{loss}^{dB}$ value also whittles down a significant portion of the per-waveguide OPB to render the remaining OPB to be available for DWDM (Fig. \ref{Fig:2}). This outcome presents the in-equality in Eq. (5) as the necessary condition to accommodate N$_\lambda$ wavelength channels in the link.


Therefore, for a given N$_\lambda$ wavelength channels in a photonic link (N$_\lambda$), total losses and power penalties experienced by these wavelength channels should be within the optical power budget as depicted in Eq. (1). It is intuitive from Fig. 2 and Eq.(1) that to design a high bandwidth photonic link, OPB should be high and, link losses and power penalties should be low. OPB can be increased by increasing the MAOP whereas power penalties in a photonic link can be reduced by increasing the FSR. Detailed discussion regarding the impact of several link design parameters on OPB and the aggregate bandwidth of a photonic link is provided in the upcoming section. 

\begin{equation}
    P_{loss}^{dB} = PP_{BER}^{dB} + IL^{dB}
\end{equation}

\begin{equation}
    OPB\ per\ Wavelength\ (dB) \geq P_{loss}^{dB}
\end{equation}

\begin{equation}
    OPB\ per\ Waveguide\ (dB) \geq P_{loss}^{dB} + 10\times log_{10}(N_\lambda)
\end{equation}

Intuitively, the bandwidth of an on-SiPhI link can be increased by increasing the (N$_\lambda$ $\times$ \textit{BR}) for the link. However, from Fig. \ref{Fig:2}, there should be sufficient OPB available for DWDM in the link to support such an increase in (N$_\lambda$ $\times$ \textit{BR}). But unfortunately, it is well established in prior works \cite{bahadori2017}\cite{bahadori2018} that (N$_\lambda$ $\times$ \textit{BR}) in the state-of-the-art on-chip and off-chip SiPh links cannot be sufficiently increased to realize $>$1 Tb/s link bandwidth, due to the low values of OPB available for DWDM that is inflicted by the current, nascent state of the SiPh technology. This unpleasant shortcoming motivated us to undertake a critical thinking exercise to identify the key design pathways towards the realization of $>$1 Tb/s on-SiPhI links. The outcomes of this exercise are presented in the next section.

\section{Identifying the Key Design Pathways Towards Multi-Terabit \\ On-SiPh-Interposer Links}
From the discussion in Section 2.2, increasing the bandwidth of an on-SiPhI link requires increasing the (N$_\lambda$ $\times$ \textit{BR}) for the link, which in turn requires a sufficient increase in the 'available OPB for DWDM' in the link. From Fig. \ref{Fig:2}, increasing the 'available OPB for DWDM' in the link can be achieved in the following ways: (i) by decreasing the total insertion loss (IL$^{dB}$) in the link; (ii) by increasing the per-waveguide MAOP for a given per-wavelength input power (Fig. \ref{Fig:2}); (iii) by decreasing the total power penalty (PP$_{BER}^{dB}$) in the link. 

The insertion loss (IL$^{dB}$) of an on-SiPhI link can be decreased by decreasing the propagation loss and coupling loss in the link (Section 2.2). Several optimization methods and fabrication processes pertaining to reducing the coupling loss in on-SiPhI links have been introduced in prior works \cite{sanchez2021,he2021,mu2020,hsu2022}. The total propagation loss in an on-SiPhI link is the product of the waveguide length (cm) and the propagation loss constant (dB/cm). Therefore, to reduce the propagation loss in an on-SiPhI link, it is intuitive that the propagation loss constant (dB/cm) should be reduced. Another way of reducing the influence of insertion loss on the bandwidth of an on-SiPhI link is to increase the per-wavelength MAOP (Fig. \ref{Fig:2}). Doing so can increase the tolerance for higher total insertion loss. Increasing the per-wavelength MAOP would in turn increase the per-waveguide MAOP. All of these factors collectively can increase the 'available OPB for DWDM'.

On the other hand, the total power penalty PP$_{BER}^{dB}$ for a link is the function of the MRR Q-factor, channel BR, Free Spectral Range (FSR), and N$_\lambda$. Prior works \cite{sai2021}, \cite{karempudi2020}, and \cite{vatsavai2020} have shown that PP$_{BER}^{dB}$ for a SiPh on-chip link can be minimized by designing the link using the optimum combination of the triplet \{MRR Q-factor, channel BR, N$_\lambda$\} for given FSR. This means that it is possible to minimize the increase in PP$_{BER}^{dB}$ caused due to the intended increase in (N$_\lambda$ $\times$ \textit{BR}) by simply employing an optimal MRR Q-factor that corresponds to the increased (N$_\lambda$ $\times$ \textit{BR}). However, precisely defining the MRR Q-factor at the design time has been proven to be very difficult due to the uncertainties emanating from the unavoidable fabrication-process non-uniformity \cite{sun2019}\cite{wang2020}. Moreover, the achievable operational bandwidth (i.e., the operating \textit{BR}) for the MRR modulator and filter devices highly depend on the utilized device fabrication process \cite{stojanovic2018}\cite{atabaki2018}. Therefore, in the wake of such dependence on the fabrication process, the more practical solution for viably increasing the bandwidth of an on-SiPhI link becomes to accept the MRR Q-factor and channel \textit{BR} that the utilized fabrication process provides, and then look to increase the N$_\lambda$ of the link. To this end, a possible, good option for lessening the impact of increasing N$_\lambda$ on PP$_{BER}^{dB}$ would be to push for as large FSR as possible, because a large FSR renders a high spectral bandwidth available for DWDM. 

Based on this discussion, we identify the following three key design pathways towards realizing $>$1 Tb/s on-SiPhI links.

\begin{itemize}
    \item \textbf{Pathway 1:} increase the available OPB for DWDM in the on-SiPhI link by minimizing the insertion losses in the link;
    \item \textbf{Pathway 2:} increase the available OPB for DWDM in the on-SiPhI link by increasing the per-wavelength and per-waveguide MAOP limits of the link;
    \item \textbf{Pathway 3:} increase the spectral bandwidth available for DWDM and minimize the power penalties in the on-SiPhI link by pushing for as large FSR as possible.
\end{itemize}


The detailed discussion on each of these pathways and the considerations made for our link-level and system-level analysis are provided in the upcoming subsections. 

\subsection{Pathway 1: Minimize Insertion Losses}
Insertion losses in an SiPh link include waveguide propagation losses and coupling losses. The amount of coupling losses incurred in on-SiPhI links depend on the utilized fabrication process for realizing waveguides and couplers. Various optimization methods and fabrication processes pertaining to reducing the coupling losses in on-SiPhI links have been introduced in prior works \cite{sanchez2021,he2021,mu2020,hsu2022}. By utilizing these, the coupling losses per on-SiPhI link can be reduced to as low as $\sim$1dB.

 Propagation losses in a silicon waveguide comprises of absorption losses and scattering losses. \textit{\textbf{Absorption losses:}} Silicon waveguides operating at wavelengths ranging from 1500-1600 nm are prone to high absorption losses due to strong two-photon absorption (TPA), despite their moderate-to-low material absorption losses in this wavelength range. This is because, for DWDM applications, when multiple wavelengths are coupled into a silicon waveguide, the total optical power in the waveguide increases which in turn induces TPA effect in the silicon waveguide \cite{hendry2014}. Due to TPA, free carrier concentration in the silicon waveguide increases that induces free-carrier absorption (FCA) \cite{hendry2014}\cite{thakkar2016} effect, which consequently increases the absorption losses in the silicon waveguide. \textit{\textbf{Scattering losses:}} Silicon waveguides are also prone to high scattering losses mainly due to the following reasons. First, sidewall roughness of the waveguides arising from fabrication imperfections. Second, high index contrast between the core (silicon) and cladding (silicon dioxide) of the waveguides. Due to the high index contrast between the core (silicon) and cladding (silicon dioxide) of a silicon waveguide, the interaction of the guided optical mode with the rough sidewalls of the waveguide increases. This enhanced mode-roughness interaction increases the scattering losses in silicon waveguides. 
 
 Therefore, high absorption and scattering losses give rise to high propagation loss in silicon waveguides, which in turn increases the amount of insertion loss present in the link. This increase in insertion loss whittles down the OPB restricting the aggregate bandwidth of photonic links. Prior works have demonstrated new photonic platforms for which TPA is absent \cite{karempudi2020}\cite{wilmart2019}\cite{rahim2017}, and such platforms can render decreased waveguide propagation losses. On the other hand, some prior works have reported silicon waveguide propagation losses below 1 dB/cm \cite{hu,dong}. The type of waveguide demonstrated in these prior works \cite{hu,dong} is a ridge waveguide, in which the interaction of guided mode with the sidewalls of the waveguide is low, thereby reducing the scattering losses. However, ridge waveguides are not compatible to couple with MRRs for cascaded DWDM. In contrast, channel waveguides are compatible for cascaded DWDM, but the lowest reported propagation loss for channel waveguides is greater than 2 dB/cm \cite{younis2018}\cite{dong2017}. 

From the above discussion, it is clear that \textit{reducing the propagation loss to 1 dB/cm and coupling loss to 1 dB} is the most optimistic goal for the near future. Therefore, we have chosen these loss values for our analysis in this paper. 

\subsection{Pathway 2: Increase Per-Wavelength and Per-Waveguide MAOP Limits}
\textbf{Per-waveguide MAOP:} As discussed in Section 2.2, the per-waveguide MAOP limit manifests in a rectilinear on-SiPhI waveguide due to the presence of very high absorption losses at relatively high optical power density and large number of multiplexed wavelength channels in the waveguide. Such high absorption losses are caused in a DWDM based silicon waveguide due to the strong two-photon absorption (TPA) and four-wave mixing nonlinearities of the silicon material in the optical C-band of operation \cite{bergman2014}\cite{lee2008}. Due to the TPA effect, the free-carrier concentration in a silicon rectilinear waveguide can dramatically increase for the input optical power densities of greater than 1 W/$\mu m^2$ (corresponds to 100 mW (20 dBm) optical power in the waveguide with the cross-sectional waveguide dimensions of 520 nm $\times$ 220 nm \cite{lee2008}), which consequently triggers free-carrier absorption (FCA) related very high propagation losses that can amount to up to 1 dB/cm additional loss per added multiplexed channel in the waveguide \cite{lee2008}. To avoid such high, power-dependent propagation losses in the waveguide, prior works limit the MAOP per waveguide to be 100 mW (20 dBm) \cite{hendry2014}\cite{wang2018}. Clearly, the introduction of the per-waveguide MAOP limit caps the available OPB for DWDM (Fig. \ref{Fig:2}), which in turn limits the achievable increase in N$_\lambda$ and link bandwidth. Therefore, we can intuitively argue that the opportunities for increasing the available OPB for DWDM can be improved by increasing, or even virtually eliminating the per-waveguide MAOP limit. Prior work \cite{ophir2010} has shown that such optical power-dependent losses are not present in silicon nitride waveguides, but due to the lack of active devices in silicon nitride material platform \cite{wilmart2019}, silicon nitride waveguides are not yet commonly used in the mainstream SiPh designs. Alternatively, another prior work \cite{karempudi2020} has shown that the per-waveguide MAOP limit can be increased, or even  be virtually eliminated, by designing SiPh links that can operate at relatively long wavelengths around 4$\mu$m. At such long wavelengths, silicon’s band gap energy is more than the energy of 2 photons, and hence, the TPA effect is absent, eliminating the optical power-dependent dramatic increase in waveguide propagation losses. Leveraging these benefits however requires adopting a new SiPh fabrication material system, referred to as silicon-on-sapphire (SOS) \cite{karempudi2020}. Although it is not clear yet if, how, and by when the SOS based SiPh designs will replace the SOI based SiPh designs, it is worth to ask this question nevertheless: Can eliminating the per-waveguide MAOP limit in on-SiPhI links boost their bandwidth beyond 1 Tb/s? To find the answer to this question, we aim to eliminate the per-waveguide MAOP, and hence, per-waveguide OPB in on-SiPhI links as part of this design pathway.

\textbf{Per-wavelength MAOP:} On the other hand, the cause for the per-wavelength MAOP limit is the interplay of the mutually conflicting free-carrier dispersion and thermal dispersion phenomena in MRR modulators that renders the modulators inoperable for per-wavelength input optical power of greater than the MAOP limit \cite{de2019}\cite{Qi2012}. Evidently, this interplay is exacerbated due to the strong TPA effect and high intra-cavity power buildup present in the silicon MRR modulators \cite{de2019}\cite{Qi2012}. Nevertheless, the MRR modulators can be intelligently designed to balance the interplay of these conflicting phenomena \cite{luo2012}, to consequently increase the per-wavelength MAOP limit to 5 mW (7 dBm) \cite{Qi2012} (which is greater than 3,2 mW (5 dBm), as commonly assumed in several link- and system-level prior works \cite{bahadori2017}\cite{wang2018}\cite{sai2021}). This outcome encourages the efforts focused on eliminating the TPA effect from MRR modulators, in hopes of further increasing the per-wavelength MAOP limit to consequently increase the per-wavelength OPB (Section 2.2; Fig. \ref{Fig:2}). However, since it is not yet clear by how much eliminating the TPA effect would impact the per-wavelength MAOP limit, we assume a relatively optimistic value of 31.5 mW (15 dBm) for the per-wavelength MAOP limit as part of this design pathway, to subsequently aim to find the answer to the following question: Can eliminating the per-wavelength MAOP limit in long on-SiPhI links (about 10 cm long) boost their bandwidth beyond 1 Tb/s?

\begin{figure}[h!]
    \centering
    \includegraphics[scale = 0.6]{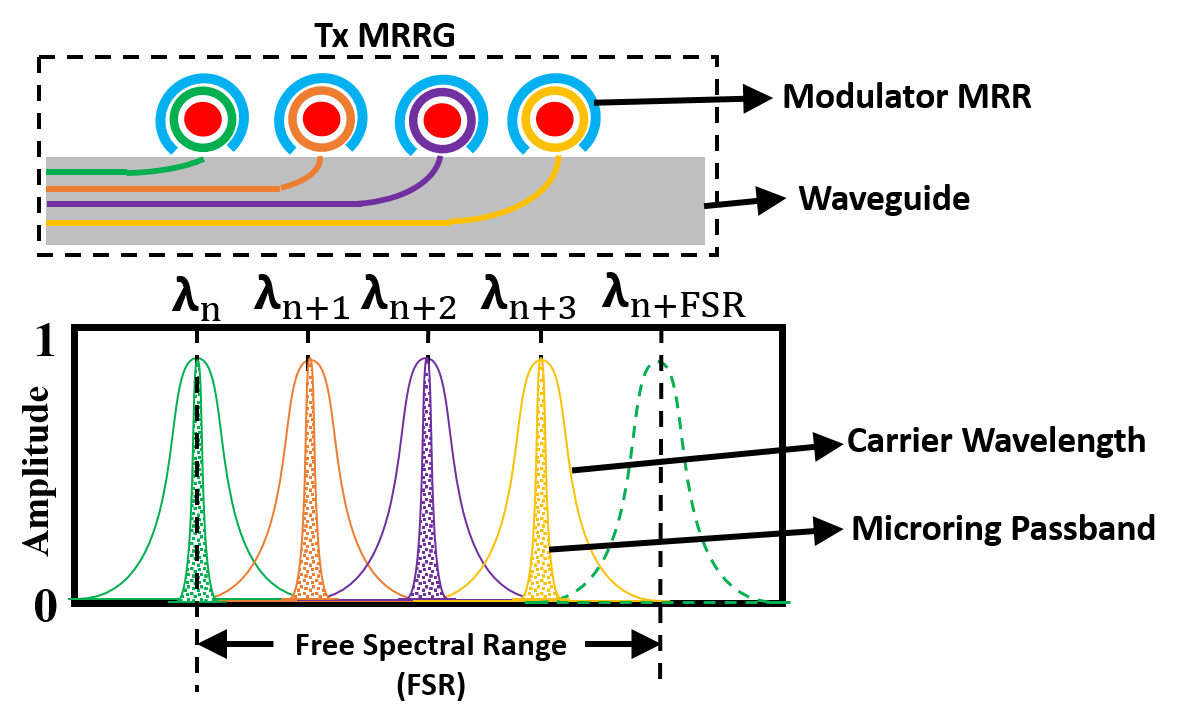}
    \caption{Illustration of FSR of an MRR.}
    \label{Fig:3}
\end{figure}

\subsection{Pathway 3: Push for as Wide FSR as Possible}
MRR, which is considered as the workhorse of a photonic link, is a looped waveguide in which the resonance occurs when the optical path length of the MRR is exactly a whole number of wavelengths. Therefore, MRRs support multiple resonances and the spacing between these resonances is FSR as shown in Fig. \ref{Fig:3}. Low values of FSR means for a given number of wavelength channels in a photonic link ($N_\lambda$), spacing between the adjacent channels is low resulting in inter-channel crosstalk \cite{bahadori2016} which in turn increases the PP$_{BER}^{dB}$ (Section 2.2). Prior works have demonstrated that low FSR of MRR devices in SOI photonic links \cite{bahadori2017} restricts the aggregate bandwidth to < 1Tb/s because of this increase in PP$_{BER}^{dB}$. Hence, it is important to enhance the FSR of constituent MRR devices to achieve the aggregate bandwidth of > 1Tb/s. 

FSR of an MRR is inversely proportional to its round-trip optical length. Therefore, to widen the FSR, one way is to reduce the round-trip optical length of the MRR which would result in a compact size of the MRR. But this length cannot be infinitely reduced due to various reasons. Firstly, reducing the round-trip optical length of an MRR increases the complexity of implementing the MRR tuning mechanism. Secondly, due to the shorter coupling length, the efficient coupling between the bus waveguide and the MRR becomes difficult to realize. Finally, reducing the round-trip optical length often results in sharper bend radius, which causes extra radiation losses and scattering losses in the MRR due to the guided optical mode that overlaps with and extends beyond the rough outer wall of the MRR bend. 

Alternatively, prior works have demonstrated various designs of MRR filters that can support larger FSR. Most recently, FSR-free MRR filter architectures were also demonstrated. Among the designs of MRR filters that support large FSR, Li Ang et al. in \cite{li2016} demonstrated a novel method that widens the FSR by means of internal reflections inside the MRR. No extra optical loss is introduced and a wide FSR up to 150 nm can be achieved using this method. Similar design has also been demonstrated in \cite{urbonas2015} that supports FSR up to 175 nm. On the other hand, FSR-free MRR filter architectures demonstrated so far in the literature are based on either integrating the contra-directional couplers (CDCs) with the MRR or by cascading MRRs with different FSRs (popularly known as vernier scheme \cite{griffel2000}). Eid. N et al. in \cite{eid2016} demonstrated FSR free MRR filters based on partially wrapping the contra-directional couplers (CDC) around the MRR. This design significantly suppresses the side-modes of the MRR resulting in FSR free response. Another similar type of FSR free MRR filter design has also been demonstrated in \cite{mistry2020} which is based on integrating the bent CDCs into the through port coupling region of the MRR cavity which suppresses all the modes except the resonance mode of the cavity. An FSR-free MRR filter architecture based on vernier scheme is demonstrated in \cite{morichetti2021}, which is polarization diverse and can be tuned beyond the range of C-band. This design of FSR-free MRR filter based on vernier scheme is CMOS compatible, making it easier to fabricate compared to other FSR-free MRR filter designs demonstrated so far. Another FSR-free MRR filter based on vernier scheme is demonstrated in \cite{petrini2021}. An FSR free MRR filter using photonic crystal cavities was also demonstrated in \cite{zhou2017}. 

Although, prior works have demonstrated MRR filters that virtually eliminate the FSR, the off-chip comb laser sources employed with on-SiPhI links, demonstrated so far \cite{gaeta2019,xue2017,rizzo2019,stern2018,kim2019}, cannot provide consistently high optical power at every wavelength for a wide range of wavelengths. Based on what is known from these prior works, comb laser sources can consistently provide > 15 dBm of optical power per wavelength for up to 80 nm range only around the C and L bands. This limitation of comb laser sources curtails the available spectral bandwidth for DWDM, which in turn has the effect of having a limited FSR because a limited FSR also curtails the available spectral bandwidth for DWDM. Therefore, we consider the widest achievable FSR in the near future to be 80 nm.

\begin{table}[h!]
\caption{Various design pathways and their corresponding optimized parameters.}
\begin{tabular}{ll}
\hline
Design Pathways   & Target Parameters                                                                                                       \\ \hline
Wide FSR           & FSR relaxed to   80nm                                                                                                        \\ 
Minimized Loss    & \begin{tabular}[c]{@{}l@{}}Reduce   waveguide propagation loss to 1 dB/cm  \\ Reduce coupling loss to 1 dB\end{tabular} \\ 
Increased MAOP       & \begin{tabular}[c]{@{}l@{}}Increase per-wavelength MAOP to 31.62 mW (15 dBm)  \\ Remove per-waveguide MAOP constraints\end{tabular} \\                           \\ \hline
\end{tabular}
\label{Table:1}
\end{table}

\subsection{Pathfinding analysis}
Table \ref{Table:1} lists the identified design pathways and their corresponding target parameters which were discussed in previous subsections. From previous subsections, it is clear that the feasible solution for viably increasing the bandwidth of an on-SiPhI link is to accept the MRR Q-factor and channel BR that the utilized fabrication process provides, and then look to increase the N$_\lambda$\ of the link. However, the Q-factor and channel BR varies across different fabrication platforms. Therefore, we consider three established SiPh fabrication platforms from prior work namely 45nm SOI CMOS \cite{stojanovic2018}, 32nm SOI CMOS \cite{stojanovic2018} and Deposited poly-Si \cite{atabaki2018} for our pathfinding analysis. Table \ref{Table:2} lists the design parameters corresponding to these fabrication platforms. The parameters listed in Table \ref{Table:2}, corresponding to each platform, do not corroborate with our intended design pathway targets (Table \ref{Table:1}). Hence, we have derived eight different variants of on-SiPhI inter-chiplet links, in which seven variants are derived based on our identified design pathways (Table \ref{Table:1}) and one variant is derived based on the parameters innate to fabrication platforms (Table \ref{Table:2}). Each of these variants are listed below:                                                               
\begin{enumerate}
    \item \textbf{Fabrication$\_$Platform$\_$Name} + \textbf{\textit{Vanilla}} - This variant utilizes innate design parameters corresponding to each of the considered fabrication platforms (Table \ref{Table:2})
    \item \textbf{Fabrication$\_$Platform$\_$Name} + \textbf{\textit{Minimized Loss}} - This variant employs innate design parameters corresponding to each platform except the insertion loss parameters, which are replaced with target parameters of our \textit{Minimized Loss} design pathway (Table \ref{Table:1}) 
    \item \textbf{Fabrication$\_$Platform$\_$Name} + \textbf{\textit{Wide FSR}} - This variant avails innate design parameters corresponding to each platform except the FSR parameter, which is replaced with target parameter of our \textit{Wide FSR} design pathway (Table \ref{Table:1})
    \item \textbf{Fabrication$\_$Platform$\_$Name} + \textbf{\textit{Increased MAOP}} - This variant utilizes innate design parameters corresponding to each platform except the MAOP parameters, which are replaced with target parameters of our \textit{Increased MAOP} design pathway (Table \ref{Table:1})
    \item \textbf{Fabrication$\_$Platform$\_$Name} + (\textbf{\textit{Minimized Loss + Wide FSR}}) - This variant employs innate design parameters corresponding to each platform except the insertion loss and FSR parameters, which are replaced with target parameters of our \textit{Minimized Loss} and \textit{Wide FSR} design pathways (Table \ref{Table:1})
    \item \textbf{Fabrication$\_$Platform$\_$Name} + (\textbf{\textit{Minimized Loss + Increased MAOP}}) - This variant avails innate design parameters corresponding to each platform but replaces the insertion loss and MAOP parameters with the target parameters of our \textit{Minimized Loss} and \textit{Increased MAOP} design pathways (Table \ref{Table:1})
    \item \textbf{Fabrication$\_$Platform$\_$Name} + (\textbf{\textit{Wide FSR + Increased MAOP}}) - This variant employs innate design parameters corresponding to each platform but replaces the wide FSR and MAOP parameters with the target parameters of our \textit{Wide FSR} and \textit{Increased MAOP} design pathways (Table \ref{Table:1})
    \item \textbf{Fabrication$\_$Platform$\_$Name} + (\textbf{\textit{Minimized Loss + Wide FSR + Increased MAOP}}) - This variant employs innate design parameters corresponding to each platform but replaces the insertion loss, wide FSR and MAOP parameters with the target parameters of our \textit{Minimized Loss}, \textit{Wide FSR} and \textit{Increased MAOP} design pathways (Table \ref{Table:1})
\end{enumerate}

Replacing the \textbf{Fabrication$\_$Platform$\_$Name} with the 45nm SOI CMOS \cite{stojanovic2018}, 32nm SOI CMOS \cite{stojanovic2018} and Deposited poly-Si \cite{atabaki2018} platforms in the above list of variants, makes it a total of twenty four variants (eight variants corresponding to each platform). Detailed link-level and system-level analysis of each of these variants is provided in upcoming sections.

\begin{table}[h!]
\caption{Design Parameters for our considered SiPh fabrication processes}
\begin{tabular}{llll}
\hline
Design   Parameters     & 45nm SOI CMOS   {}\cite{stojanovic2018}{} & 32nm SOI CMOS   {}\cite{stojanovic2018}{} & Deposited   Poly-Si {}\cite{atabaki2018}{} \\ \hline
Modulator MRRs   Q       & 10000                    & 6000                     & 5000                         \\ 
Filter MRRs Q            & 8500                     & 6500                     & 5000                         \\ 
MRR Radius               & 5 µm                     & 5 µm                     & 7.5 µm                       \\ 
Operating-wavelength             & 1290 nm                  & 1310 nm                  & 1300 nm                      \\ 
FSR                     & 12.6 nm                  & 13 nm                    & 8.54 nm                      \\ 
Modulator   Bandwidth   & 13 GHz                   & 13.5 GHz                 & 16.8 GHz                     \\ 
Detector   Bandwidth    & 5 GHz                    & 12.5 GHz                 & 11 GHz                       \\ 
Sensitivity   (dBm)     & -17.645                  & -11.79                   & -20.414                      \\ 
Propagation   Loss      & 3.7 dB/cm                & 10 dB/cm                 & 20 dB/cm                     \\ 
MAOP (per-wavelength)            & 1.7 mW (2.3 dBm)                  & 2.5 mW (4 dBm)                    & 2.8 mW (4.5 dBm)                      \\ 
MAOP (per-waveguide)           & 100 mW(20 dBm)                   & 100 mW (20 dBm)                   & 100 mW (20 dBm)                       \\ 
Per-coupler   Loss      & 1.5 dB                   & 4.9 dB                   & 5.2 dB                       \\ 
Bit-rate                & 12 Gb/s                  & 12.5 Gb/s                & 11 Gb/s                      \\ 
Per-wavelength   Budget & 19.945 dB                & 15.794 dB                & 24.914 dB                    \\ 
Per-waveguide   Budget  & 37.645 dB                & 31.79 dB                 & 40.414 dB                    \\ 
Waveguide   Length      & 1-10 cm                  & 1-10 cm                  & 1-10 cm                      \\ 
Modulator IL            & 4.7 dB                    & 2.8 dB  
     & 3.8 dB \\
Filter IL               & 0.18 dB                   & 0.14 dB
    & 0.11 dB \\
Coupling   Loss   & 1.5 dB                   & 4.9 dB                   & 5.2 dB                       \\ \hline
\end{tabular}
\label{Table:2}
\end{table}

\section{Link-level evaluation}
\subsection{Evaluation Setup}
To perform the pathfinding link-level analysis for each of the 24 derived variants (Section 3.4), we utilize a search heuristic based optimization framework provided in \cite{thakkar2017}. This search heuristic consists of an error function that takes different values of $N_{\lambda}$ and channel BR as input and evaluates an error value for each duplet of ($N_{\lambda}$, BR), for a given waveguide link length. From that, the duplet of ($N_{\lambda}$, BR) corresponding to minimum positive value of error function is chosen as the optimal duplet since minimum positive value of error-function means the available OPB has been utilized to its maximum while satisfying the condition given in Eq. (5). With the obtained ($N_{\lambda}$, BR) duplet for each derived variant, we have calculated corresponding aggregate bandwidth which is the product of  $N_{\lambda}$ and channel BR, and energy per bit (EPB) which is sum of link laser power, thermal tuning power, modulator driver power and receiver power \cite{basak2008high}. The results of this analysis and a detailed discussion is provided in the next subsection.

\subsection{Results and Comparison}
\begin{figure}
    \centering
    \includegraphics[scale = 0.08]{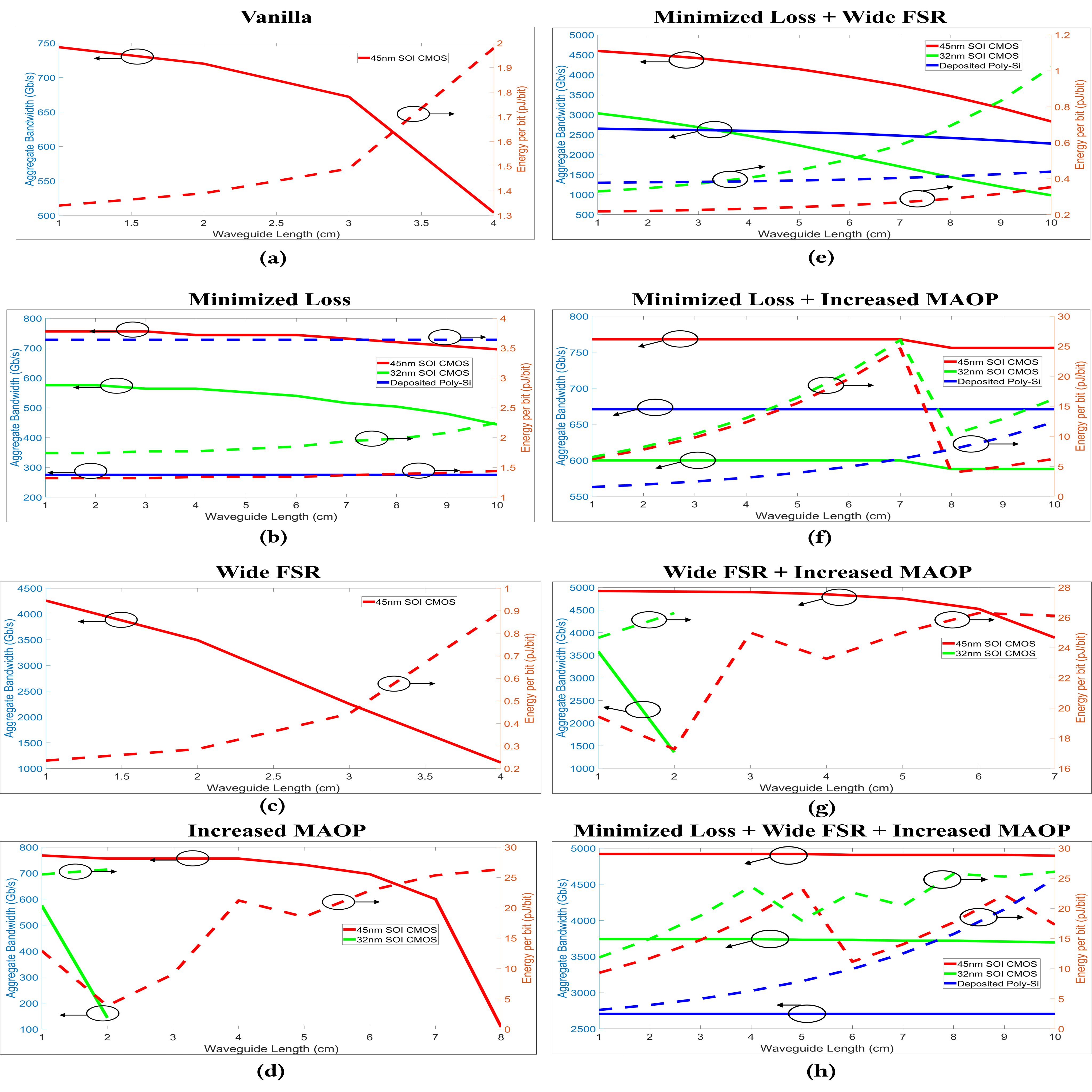}
    \caption{Aggregate bandwidth and energy per bit (EPB) values for different waveguide lengths ranging from 1 cm to 10 cm obtained from the analysis performed on (a) Vanilla, (b) Minimized Loss, (c) Wide FSR, (d) Increased MAOP, (e) Minimzed loss + Wide FSR, (f) Minimzed loss + Increased MAOP, (g) Wide FSR + Increased MAOP, and (h) Minimized Loss + Increased MAOP + Wide FSR on-SiPh variants derived from 45nm SOI CMOS \cite{stojanovic2018}, 32nm SOI CMOS \cite{stojanovic2018} and deposited poly-Si \cite{atabaki2018} platforms.}
    \label{Fig:4}
\end{figure}

Fig. \ref{Fig:4} illustrates the evaluated aggregate bandwidth (primary Y-axis) and EPB (secondary Y-axis) for different on-SiPhI inter-chiplet variants corresponding to three different SiPh fabrication platforms namely 45nm SOI CMOS \cite{stojanovic2018}, 32nm SOI CMOS \cite{stojanovic2018} and deposited poly-Si \cite{atabaki2018}, for different waveguide lengths ranging from 1 cm to 10 cm (X-axis). Based on the results obtained from this analysis, we have categorized the derived variants in to two types namely non-viable and viable variants. Non-viable variants are the variants that are unfeasible to implement as on-SiPhI inter-chiplet links due to the high insertion losses at longer waveguide lengths that exceed the amount of available OPB in the link, thereby not supporting any wavelength channels in the link and yielding no aggregate bandwidth. On the other hand, viable variants are the variants that are feasible to implement as on-SiPhI inter-chiplet links since they support some tangible aggregate bandwidth for waveguide link lengths of up to 10 cm. Detailed discussion on each category of variants is provided in next subsections.

\subsubsection{Non-Viable Variants}
Among the derived variants, \textit{Vanilla} (Fig. \ref{Fig:4}(a)), \textit{Wide FSR} (Fig. \ref{Fig:4}(c)), \textit{Increased MAOP} (Fig. \ref{Fig:4}(d)) and \textit{Wide FSR + Increased MAOP} (Fig. \ref{Fig:4}(g)) variants are considered as non-viable variants because they support no wavelength channels and therefore do not support aggregate bandwidth for longer waveguide lengths. 

Among the non-viable variants, \textit{Vanilla} variants corresponding to 32nm SOI CMOS and deposited poly-Si platforms do not support any wavelength channels due to high waveguide propagation loss of ~10 dB/cm and ~20 dB/cm respectively (Table \ref{Table:2}) resulting in excess amount of insertion loss in the link. But the \textit{Vanilla} variant corresponding to 45nm SOI CMOS platform can support wavelengths up to a waveguide length of 4 cm due to low insertion loss (3.7 dB/cm (Table \ref{Table:2})) compared to the other \textit{Vanilla} variants. However, the aggregate bandwidth and EPB of this variant is limited to 744 Gb/s and 1.34 pJ/bit respectively. Therefore, it is intuitive that reducing the insertion loss is vital in realizing longer on-SiPhI inter-chiplet links.

Similarly, \textit{Wide FSR} variants corresponding to 32nm SOI CMOS and deposited poly-Si platforms do not support any wavelength channels whereas \textit{Wide FSR} variant corresponding to 45nm SOI CMOS platform can support wavelength channels up to a waveguide length of 4 cm wwith peak aggregate bandwidth of 4.3 Tb/s and corresponding EPB of 0.235 pJ/bit, and a minimum aggregate bandwidth of 1.12 Tb/s with corresponding EPB of 0.896 pJ/bit. Therefore, it is intuitive that Widening the FSR will increase the spacing between the wavelength channels in the link resulting in low power penalty in the link and thereby increasing the available OPB for DWDM. However, the presence of high insertion losses in the link is still an impediment in realizing longer on-SiPhI inter-chiplet links. Therefore, it is lucid that implementing the \textit{Wide FSR} design pathway in combination with \textit{Minimized Loss} design pathway will aid in realizing longer on-SiPh links with superior aggregate bandwidth and energy efficiency. 

Also, \textit{Increased MAOP} variant corresponding to deposited poly-Si platform does not support any aggregate bandwidth whereas the same variant corresponding to 45nm SOI CMOS and 32nm SOI CMOS platforms can realize on-SiPhI links up to a waveguide length of 8 cm and 2 cm respectively. In terms of aggregate bandwidth and energy efficiency, \textit{Increased MAOP} variant corresponding to 45nm SOI CMOS platform achieve peak aggregate bandwidth of 768 Gb/s with corresponding EPB of 12.9 pJ/bit and a minimum aggregate bandwidth of 108 Gb/s with corresponding EPB of 26.34 pJ/bit whereas the same variant corresponding to 32nm SOI CMOS platform achieves peak aggregate bandwidth of 576 Gb/s with corresponding EPB of 25.51 pJ/bit and least aggregate bandwidth of 144 Gb/s with corresponding EPB of 26.32 pJ/bit. Therefore, it is intuitive that implementing the \textit{Increased MAOP} design pathway in combination with any other design pathways, especially the \textit{Minimized Loss} design pathway, will enable these variants to realize longer on-SiPhI links with higher aggregate bandwidth and energy efficiency.

\textit{Wide FSR + Increased MAOP} variant corresponding to 45nm SOI CMOS and 32nm SOI CMOS platforms can realize on-SiPhI links up to waveguide length of 7 cm and 2 cm respectively whereas the same variant corresponding to deposited poly-Si platform does not support any wavelength channels. In terms of performance, \textit{Wide FSR + Increased MAOP} variant corresponding to 45nm SOI CMOS platform achieves peak aggregate bandwidth of 4.92 Tb/s with corresponding EPB of 19.44 pJ/bit and a minimum aggregate bandwidth of 3.88 Tb/s with corresponding EPB of 26.13 pJ/bit whereas the same variant corresponding to 32nm SOI CMOS platform achieves peak aggregate bandwidth of 3.6 Tb/s with corresponding EPB of 24.7 pJ/bit and a minimum aggregate bandwidth of 1.4 Tb/s with corresponding EPB of 26.3 pJ/bit. Clearly, multi Tb/s aggregate bandwidth can be achieved by widening the FSR in combination with increasing the MAOP but the presence of high insertion loss in the link makes it unfeasible to realize longer on-SiPhI inter-chiplet links. 

Therefore, implementing the \textit{Minimized Loss} design pathway in combination with other design pathways is the key to realizing longer on-SiPhI inter-chiplet links with >1Tb/s aggregate bandwidth and <1pJ/bit energy efficiency. In addition, using repeaters can also make \textit{Vanilla}, \textit{Wide FSR} and \textit{Wide FSR + Increased MAOP} variants corresponding to 45nm SOI CMOS platform, and \textit{Wide FSR + Increased MAOP} variant corresponding to 32nm SOI CMOS platform viable for longer waveguide lengths.

\subsubsection{Viable Variants}
Among the derived variants, \textit{Minimized Loss} (Fig. \ref{Fig:4}(b)), \textit{Minimized Loss + Wide FSR} (Fig. \ref{Fig:4}(e)), \textit{Minimized Loss + Increased MAOP} (Fig. \ref{Fig:4}(f)) and \textit{Minimized Loss + Wide FSR + Increased MAOP} (Fig. \ref{Fig:4}(h)) variants are considered as viable variants to implement on-SiPhI inter-chiplet links since they support wavelength channels upto waveguide length as long as 10 cm. 

Among these viable variants, \textit{Minimized Loss} variant corresponding to 45nm SOI CMOS platform achieves peak aggregate bandwidth of 756 Gb/s with corresponding EPB of 1.32 pJ/bit and a minimum aggregate bandwidth of 696 Gb/s with corresponding EPB of 1.4 pJ/bit whereas the same variant corresponding to 32nm SOI CMOS platform acheives peak aggregate bandwidth of 576 Gb/s with corresponding EPB of 1.74 pJ/bit and a minimum aggregate bandwidth of 444 Gb/s with corresponding EPB of 2.25 pJ/bit. Similarly, \textit{Minimized Loss} variant corresponding to deposited poly-Si platform supports peak aggregate bandwidth of 275 Gb/s with corresponding EPB of 3.64 pJ/bit. Hence, it is evident that minimizing the insertion loss in the link will enable the variants to realize on-SiPhI links up to a waveguide length as long as 10 cm. However, these variants do not acheive aggregate bandwidth of more than 1Tb/s and EPB of less than 1 pJ/bit due to high power penalty in the link resulting from the low FSR of the considered SiPh fabrication platforms and also due to low MAOP in the link resulting in less available OPB. Therefore, minimzing the insertion loss in combination with other design pathways is vital to yield extremely high aggregate bandwidth and energy-efficient on-SiPhI inter-chiplet links which is the most important step towards enabling the chiplet based systems for the future.  

As illustrated in Fig. \ref{Fig:4}, \textit{Minimized Loss + Increased MAOP} variant corresponding to 45nm SOI CMOS platform achieves peak aggregate bandwidth of 768 Gb/s with corresponding EPB of 6.18 pJ/bit and a minimum aggregate bandwidth of 756 Gb/s with corresponding EPB of 6.26 pJ/bit whereas the same variant corresponding to 32nm SOI CMOS platform acheives peak aggregate bandwidth of 600 Gb/s with corresponding EPB of 6.54 pJ/bit and a minimum aggregate bandwidth of 588 Gb/s with corresponding EPB of 16.25 pJ/bit. Similarly, \textit{Minimized Loss + Increased MAOP} variant corresponding to deposited poly-Si platform yields peak aggregate bandwidth of 671 Gb/s with corresponding EPB of 1.56 pJ/bit. Here, minimizing the loss in combination with increasing the MAOP will enable the variants to support higher aggregate bandwidth compared to the \textit{Minimized Loss} variants discussed previously. But the amount of power penalty in the link is still high and increase in MAOP along with reducing the insertion loss does not offset that penalty. Therefore, it is lucid from the observations that minimizing the loss in combination with enhancing the FSR or implementing all three design pathways in combination will enable the on-SiPhI inter-chiplet links to achieve higher aggregate bandwidth and energy-efficiency.

As we can depict from Fig. \ref{Fig:4}(e) and Fig. \ref{Fig:4}(h), \textit{Minimized Loss + Wide FSR} and \textit{Minimized Loss + Wide FSR + Increased MAOP} variants achieve more than 1Tb/s aggregate bandwidth up to waveguide length of 10cm, for all the considered SiPh fabrication platforms, in which former variant corresponding to 45nm SOI CMOS platform achieves peak aggregate bandwidth of 4.6 Tb/s with corresponding EPB of 0.218 pJ/bit whereas the latter variant corresponding to the same fabrication platform achieves peak aggregate bandwidth of 4.92 Tb/s with corresponding EPB of 9.2 pJ/bit. 

Comparing the evaluated aggregate bandwidth and EPB obtained from the analysis performed on the derived on-SiPhI inter-chiplet variants, we can deduce that for different waveguide lengths ranging from 1 cm to 10cm, \textit{Minimized Loss + Wide FSR + Increased MAOP} variant corresponding to 45nm SOI CMOS platform achieves highest aggregate bandwidth whereas Minimized Loss + Wide FSR variant corresponding to the same platform acheives lowest EPB. The \textit{Minimized Loss + Wide FSR} variant corresponding to 45nm SOI CMOS has low crosstalk and signal truncation penalty due to high FSR, and optimum values of modulator and detector Q. Therefore, due to low insertion loss and low power penalty, \textit{Minimized Loss + Wide FSR} variant corresponding to 45nm SOI CMOS platform achieves lowest EPB among all the variants for waveguide lengths up to 10 cm. But this variant falls short of achieving highest aggregate bandwidth due to low OPB in the link resulting from low MAOP. On the other hand, \textit{Minimized Loss + Wide FSR + Increased MAOP} variant corresponding to 45nm SOI CMOS platform also has low insertion loss and increasing the MAOP per-wavelength enables this variant to accommodate higher number of wavelength channels which in turns enables it to achieve highest aggregate bandwidth among all the variants. However, higher number of wavelength channels in the link i.e., the high degree of DWDM leads to less channel spacing which in turn will increase the crosstalk penalty in the link resulting in higher EPB consumption. Therefore, designing on-SiPhI inter-chiplet links by implementing all three design pathways using 45nm SOI CMOS platform can achieve peak aggregate bandwidth of 4.92 Tb/s whereas EPB <1 pJ/bit with corresponding aggregate bandwidth of 4.6 Tb/s can be achieved by implementing the \textit{Minimized Loss} design pathway in combination with \textit{Wide FSR} for the same fabrication platform.  

From the above observations, we can notice that in order to design longer on-SiPhI inter-chiplet links for the future, it is vital to keep the insertion loss to minimum. Similarly, we can also infer that combining the other design pathways such as \textit{Wide FSR} and \textit{Increased MAOP} with \textit{Minimized Loss} can scale the aggregate bandwidth to more than 1Tb/s which is the most important step towards meeting the bandwidth requirements of future chiplet-based computing systems. However, it is important to see how these variants perform at system-level. Therefore, we perform a system-level analysis by implementing the derived variants on a CPU based multi-core multi-chiplet architecture and a GPU based multi-chiplet module (MCM) considered from prior work (\cite{bashir2017}\cite{khani2021}). Details of this analysis are provided in the next section.

\begin{table}[]
\caption{Inter-chiplet variants derived from 45nm SOI CMOS, 32nm SOI CMOS and Deposited poly-Si platforms.}
VR: Viable with repeaters, V: Viable, NV: Non-viable\footnote{}
\begin{tabular}{cccccccccc}
\hline
Variants                                                                & \multicolumn{3}{c}{\begin{tabular}[c]{@{}c@{}}45nm SOI \\ CMOS\end{tabular}} & \multicolumn{3}{c}{\begin{tabular}[c]{@{}c@{}}32nm SOI \\ CMOS\end{tabular}} & \multicolumn{3}{c}{\begin{tabular}[c]{@{}c@{}}Deposited \\ Poly-Si\end{tabular}} \\ \cline{2-10} 
                                                                                         &      & (N$_\lambda$,BR)     & \begin{tabular}[c]{@{}c@{}}ADR \\ (Gb/s)\end{tabular}   &      & (N$_\lambda$,BR)     & \begin{tabular}[c]{@{}c@{}}ADR \\ (Gb/s)\end{tabular}   &        & (N$_\lambda$,BR)      & \begin{tabular}[c]{@{}c@{}}ADR \\ (Gb/s)\end{tabular}    \\ \hline
Vanilla                                                                                  & VR   & (42, 12)    & 504                                                     & NV   &             &                                                         & NV     &              &                                                          \\ 
Minimized   Loss                                                                         & V    & (60, 12)    & 720                                                     & V    & (42, 12)    & 504                                                     & V      & (25, 11)     & 275                                                      \\ 
Wide FSR                                                                                 & VR   & (93, 12)    & 1116                                                    & NV   &             &                                                         & NV     &              &                                                          \\ 
Increased MAOP                                                                           & V    & (9, 12)     & 108                                                     & VR   & (12, 12)    & 144                                                     & NV     &              &                                                          \\ 
\begin{tabular}[c]{@{}c@{}}Minimized   Loss + \\ Wide FSR\end{tabular}                   & V    & (289, 12)   & 3468                                                    & V    & (120, 12)   & 1440                                                    & V      & (220, 11)    & 2420                                                     \\ 
\begin{tabular}[c]{@{}c@{}}Minimized   Loss +\\  Increased MAOP\end{tabular}               & V    & (63, 12)    & 756                                                     & V    & (49, 12)    & 588                                                     & V      & (61, 11)     & 671                                                      \\ 
\begin{tabular}[c]{@{}c@{}}Wide FSR   +\\  Increased MAOP\end{tabular}                     & VR   & (404, 12)   & 4848                                                    & VR   & (113, 12)   & 1356                                                    & NV     &              &                                                          \\
\begin{tabular}[c]{@{}c@{}}Minimized   Loss +\\  Wide FSR +\\  Increased MAOP\end{tabular} & V    & (409, 12)   & 4908                                                    & V    & (310, 12)   & 3720                                                    & V      & (246, 11)    & 2706                                                     \\ \hline
\end{tabular}
\label{Table:3}
\end{table}

\section{System-Level Evaluation}
\subsection{CPU based multi-core multi-chiplet architecture}
We have performed system-level analysis on a CPU based multi-core multi-chiplet architecture named NUPLet \cite{bashir2017} and on a GPU based MCM from \cite{khani2021}. The architecture, inter-chiplet network of the NUPLet and the design of GPU based MCM are described in following subsections.

\begin{figure}
    \centering
    \includegraphics[scale = 0.1]{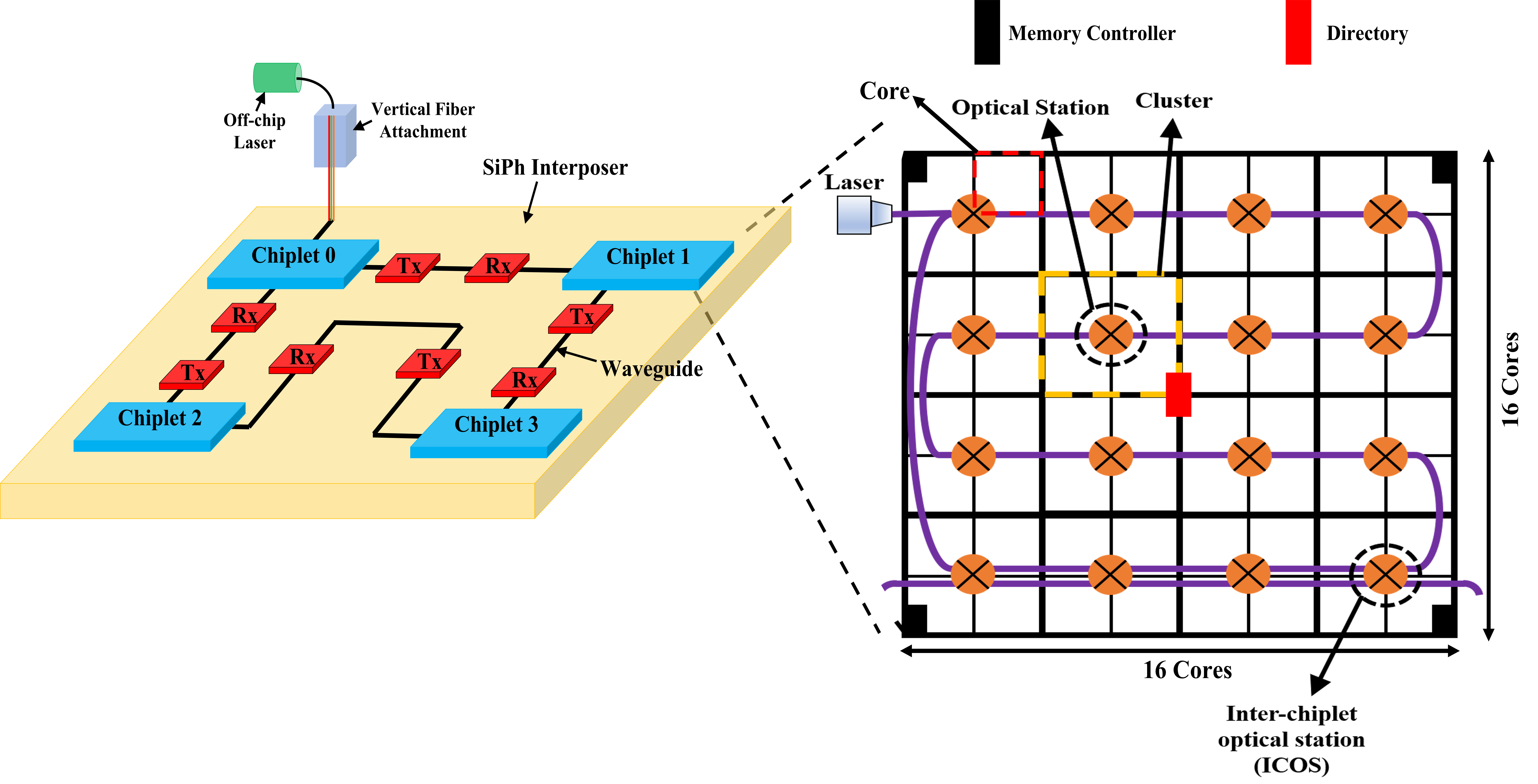}
    \caption{Chiplet based Design of NUPLet.}
    \label{Fig:5}
\end{figure}

\begin{figure}
    \centering
    \includegraphics[scale = 0.075]{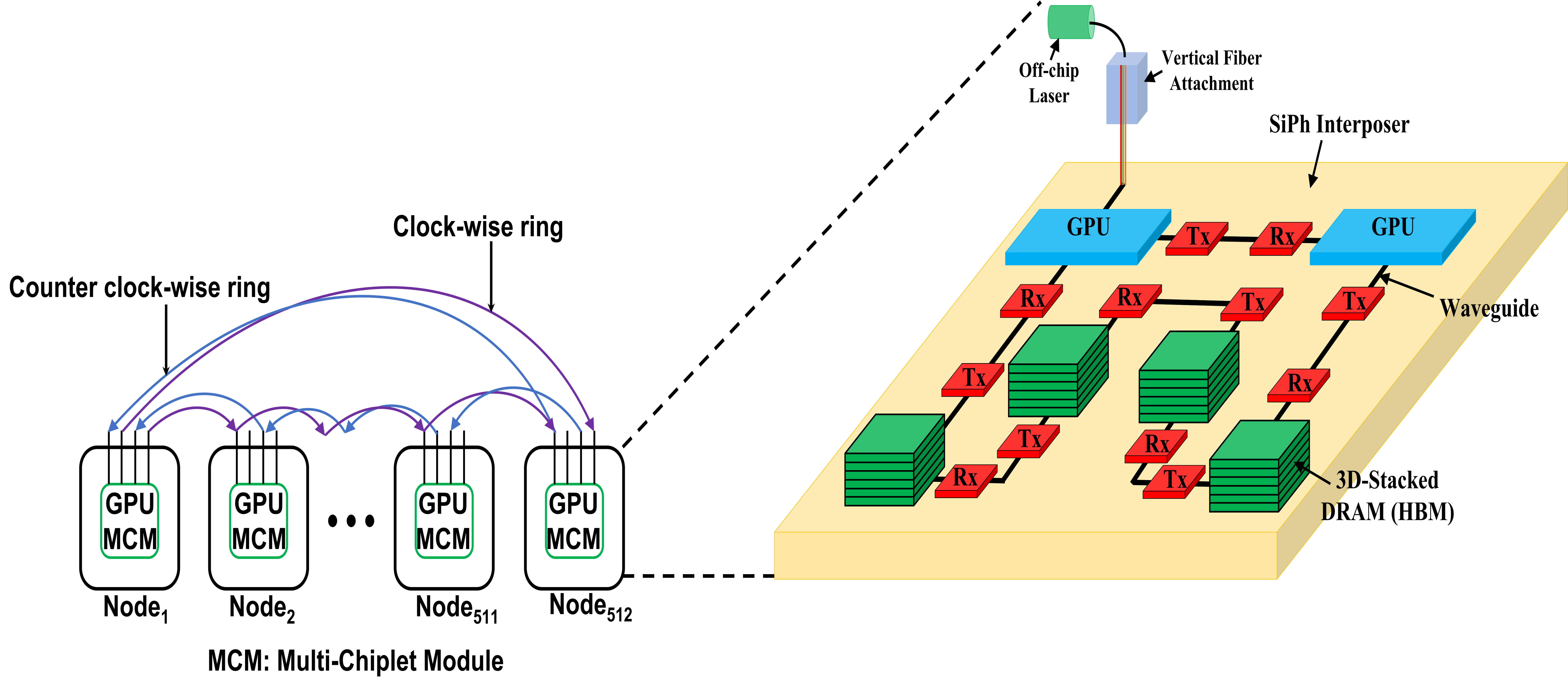}
    \caption{GPU based multi-chiplet module (MCM).}
    \label{Fig:6}
\end{figure}

\subsubsection{Architecture of NUPLet}
NUPLet architecture (Fig. \ref{Fig:5}) consists of four chips and each chip is called a chiplet. Each chiplet is composed of 32 cores divided into 16 clusters with 2 cores per cluster. Each chiplet in NUPLet also has an 8MB last level cache (LLC) divided into 32 cache banks in 16 clusters with 2 cache banks per cluster. At the interface of each cluster in a chiplet, an optical station is present which consists of a transmitter (modulator MRs array) and a receiver (filter MRs array) that enable inter-chiplet and intra-chiplet data communication. SOI based waveguides connect optical stations in a chiplet with one another in a crossbar configuration. Each optical station in NUPLet receives some amount of multi-wavelength optical power through waveguides via an off-chip laser that can generate up to 180mW of optical power. Whenever an optical station wants to send data, it redirects some portion of the light from the waveguide. This light is split into multiple wavelengths using a comb splitter. The electrical data packet from the core is converted to parallel electrical data signals and modulated onto these wavelengths using modulator MRs. These modulated wavelengths travel along the waveguide to the destination station where a bank of MRR filters drop these wavelengths onto the adjacent photodetectors to regenerate the electrical data signals and consequently, the electrical data packet which is passed onto the destination core. Intra-chiplet network in NUPLet is based on SWMR (single writer multiple reader) crossbar topology \cite{pan2009} where each optical station is connected to the other optical stations in chiplet using a dedicated waveguide. Similarly, Inter-chiplet network in NUPLet is based on MWMR (multiple writer multiple reader) crossbar topology \cite{pan2010}. Detailed discussion on inter-chiplet network is provided in the upcoming subsection. 

\subsubsection{Inter-Chiplet Network of NUPLet}
Optical stations at the bottom of each chiplet are used for both intra-chiplet and inter-chiplet communication and are called as inter-chiplet optical stations (ICOS) as shown in Fig. \ref{Fig:5}. There are a total of 16  ICOSs with 4 ICOSs per chiplet. These ICOSs utilize MWMR crossbar topology where multiple optical stations can send and receive data using their corresponding modulator and filter MRR banks respectively which enables the stations to share the available optical bandwidth. Each ICOS also consists of queues that hold intra-chiplet and inter-chiplet messages. The inter-chiplet network of NUPLet has 8 data waveguides and 8 power waveguides. If an ICOS wishes to send data, firstly it should get access to a data-power waveguide pair, then redirect some portion of light signal from the power waveguide, use comb splitter to split the light into multiple wavelength signals, modulate the electrical data onto these wavelength channels and send it to the destination station through the data waveguide. 

The power required to transmit data or an inter-chiplet message from one chiplet to other is high compared to power required for intra-chiplet communication. This is because of longer lengths and high propagation losses of inter-chiplet waveguides. In addition, there are other insertion losses such as coupler loss, splitter loss and through loss of MRs. All of these losses increase the laser power consumption and degrade the performance. In order to minimize the laser power consumption in inter-chiplet communication, NUPLet utilizes NUCA (non-uniform cache access schemes) and a unique prediction scheme.

A miss in L1 level cache prompts a request to one of the cache banks in LLC. Cache bank that contains the block of data may lie in same chiplet from which the request was prompted or in any other chiplets. If the cache bank lies in same chiplet, then it is called home bank. Otherwise, it is called non-home bank. Analysis provided in \cite{bashir2017} shows that  57\% of these prompted requests are sent to non-home banks and only 7\% of these result in a hit. For a lower hit rate, large number of inter-chiplet messages are sent resulting in high laser power consumption. Restricting the access requests to local cache banks will reduce the number of inter-chiplet messages that can alleviate this drawback. For that, NUPLet utilizes NUCA schemes which enables the migration of requested cache block to cache banks that are on the same chiplet as the requesting cores. This will increase the hit rate and reduces the amount of inter-chiplet messages. 

Execution time of an application is divided into several fixed size durations called epochs. Several prior works have demonstrated power reduction by predicting the traffic for the next epoch by analyzing the behavior of application in previous epochs. NUPLet utilizes a similar type of prediction scheme that predicts the number of inter-chiplet and intra-chiplet messages that will be sent in the next epoch and the consequent laser power required. Accurate prediction of inter-chiplet and intra-chiplet messages will reduce the wastage of laser power and enhances the performance.

NUCA and prediction schemes of the NUPLet reduce the laser power consumption but insertion loss and power penalties in photonic links of NUPLet are still present that will result in significant amount of laser power consumption. Therefore, we implement our derived inter-chiplet variants on NUPLet architecture and perform a system-level analysis from which we evaluate performance and energy consumption. Details of this evaluation are provided in the upcoming subsections. 

\subsection{GPU based Multi-Chiplet Module}
The computation requirements of modern data centric applications such as machine learning has been partially met by swift development of hardware accelerators. Although hardware accelerators have provided a notable amount of speedup but training conventional ML models can still take a significant amount of time. Several solutions have been introduced that enable distributed training on a small number of GPUs connected with a high speed electrical switch with a Tb/s bandwidth. But future ML training workloads require several Tb/s of bandwidth per device at large scales in order to reduce the training time. This raises the need for >1 Tb/s interconnects for distributed ML systems which is implausible to achieve from conventional electrical interconnects. Therefore, in \cite{khani2021}, khani et al. proposed an end-to-end optical solution called SiP-ML for scaling of ML workloads by leveraging silicon photonic chiplets. As a part of this work, khani et al. explored two all optical architectures for scaling of ML workloads and one among them is SiP-ring shown in Fig. \ref{Fig:6}. This SiP-ring architecture consists of disaggregated GPU MCMs and the inter-chiplet communication in each of these modules occurs in photonic domain. Each of these GPU MCMs are connected to each other in a ring topology which enables communication in both directions and is easily reconfigurable. Inside each of the GPU MCMs, there are two GPUs connected to four 3D stacked DRAMs as shown in inset of Fig. \ref{Fig:6}. As a part of our system-level analysis, we implemented our derived on-SiPhI inter-chiplet variants on GPU MCMs and evaluated the impact of aggregate bandwidth of the inter-chiplet variants on the training time of conventional deep neural network (DNN) models, which are widely used in computer vision and natural language processing applications. More details of this evaluation are provided in further subsections.

\begin{figure}
    \centering
    \includegraphics[scale = 0.168]{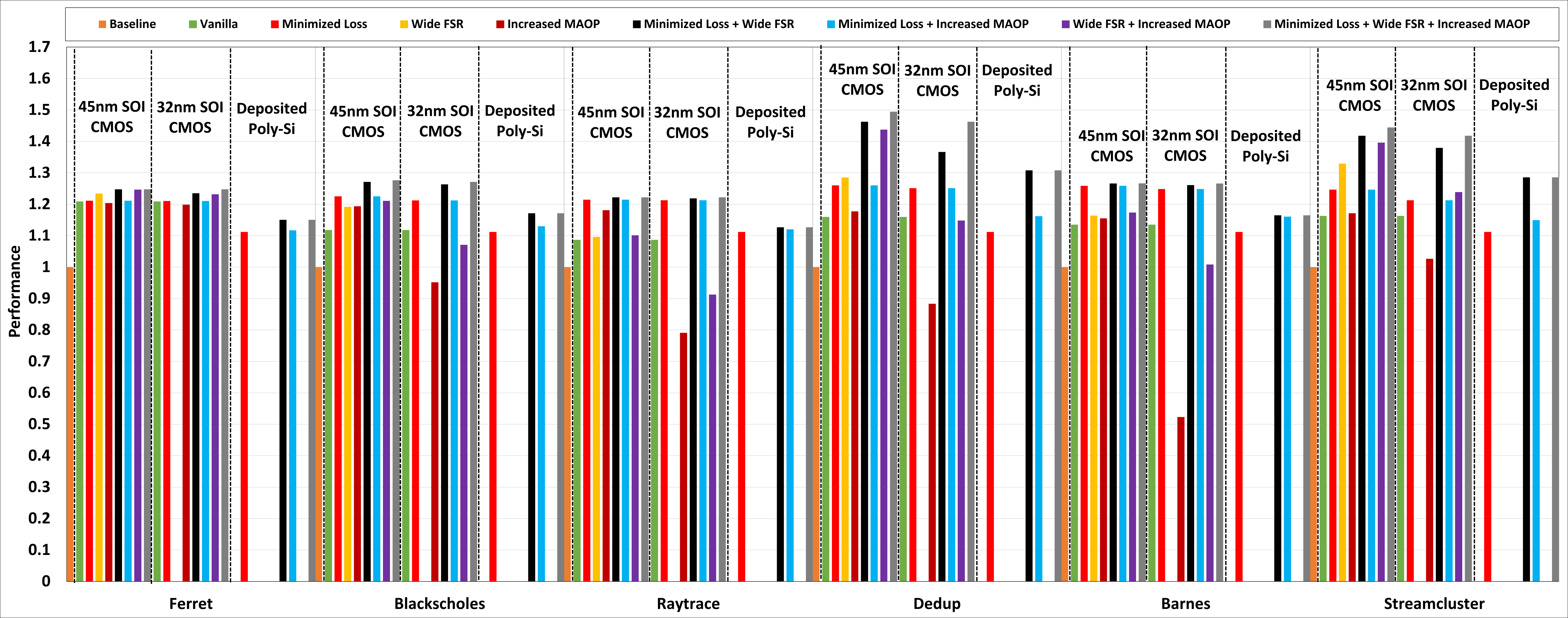}
    \caption{Performance comparison of on-SiPh variants derived from 45nm SOI CMOS, 32nm SOI CMOS and deposited poly-si photonic platforms implemented on NUPLet architecture.}
    \label{Fig:7}
\end{figure}

\subsection{Evaluation setup}
\subsubsection{CPU based multi-core multi-chiplet architecture}
As a part of our system-level analysis, we have implemented the derived on-SiPhI inter-chiplet variants (Table \ref{Table:3}) on a CPU based multi-core multi-chiplet architecture named NUPLet \cite{bashir2017} and performed a benchmark-driven simulation based analysis from which we have evaluated the performance (1/execution time), energy consumption and energy-delay product of the NUPLet architecture. We have used four 32-core chiplets in all our designs. We have evaluated our designs on a cycle architectural simulator named \textit{Tejas} \cite{sarangi2015} for real world traffic applications in the PARSEC benchmark suite \cite{bienia2008}. For all our experiments, we have used an epoch size of 100 cycles.

\subsubsection{GPU based multi-chiplet module}
For the system-level analysis on GPU based MCM \cite{khani2021}, we have utilized a simulator named \textit{Rostam} from \cite{khani2021} which is available online at \url{https://github.com/MLNetwork/rostam.git}. We implement our derived on-SiPhI inter-chiplet variants on \textit{SiP-ring} architecutre and evaluate the impact of aggregate bandwidth of inter-chiplet variants on the \textit{time-to-accuracy} of the conventional DNN models. For this analysis, we have considered three representative DNN models namely \textit{ResNet50} \cite{he2016deep}, \textit{Transformer} and \textit{Megatron} \cite{shoeybi2019megatron}. Among these models, \textit{ResNet} is an image classification model with 25 million parameters. Silimarly, \textit{Transformer} is a model with 350 million parameters whereas \textit{Megatron} is a model with 18 billion parameters. We evaluate \textit{time-to-accuracy} metric corresponding to the inter-chiplet variants implemented on the \textit{SiP-Ring} architecture for each DNN model by multiplying the time for a single iteration (obtained from the simulator) by the number of training iterations (considered from prior work \cite{shallue2018measuring}) required to reach the target accuracy.

\subsection{Evaluation Results}
For the system-level analysis, we have implemented the derived on-SiPhI inter-chiplet variants (Table \ref{Table:3}) on a CPU based multi-core multi-chiplet architecture named NUPLet \cite{bashir2017} and on a GPU based MCM from \cite{khani2021} which is used for distributed ML training. On NUPLet, we have performed a benchmark-driven simulation based analysis from which we have evaluated performance (1/execution time) and energy consumption of the NUPLet architecture. On the GPU based MCM considered from \cite{khani2021}, we have evaluated the impact of link-level aggregate bandwidth of our derived on-SiPhI inter-chiplet variants on training time of conventional ML models. The results of this analysis are discussed in the next subsection. 

\begin{figure}
    \centering
    \includegraphics[scale = 0.168]{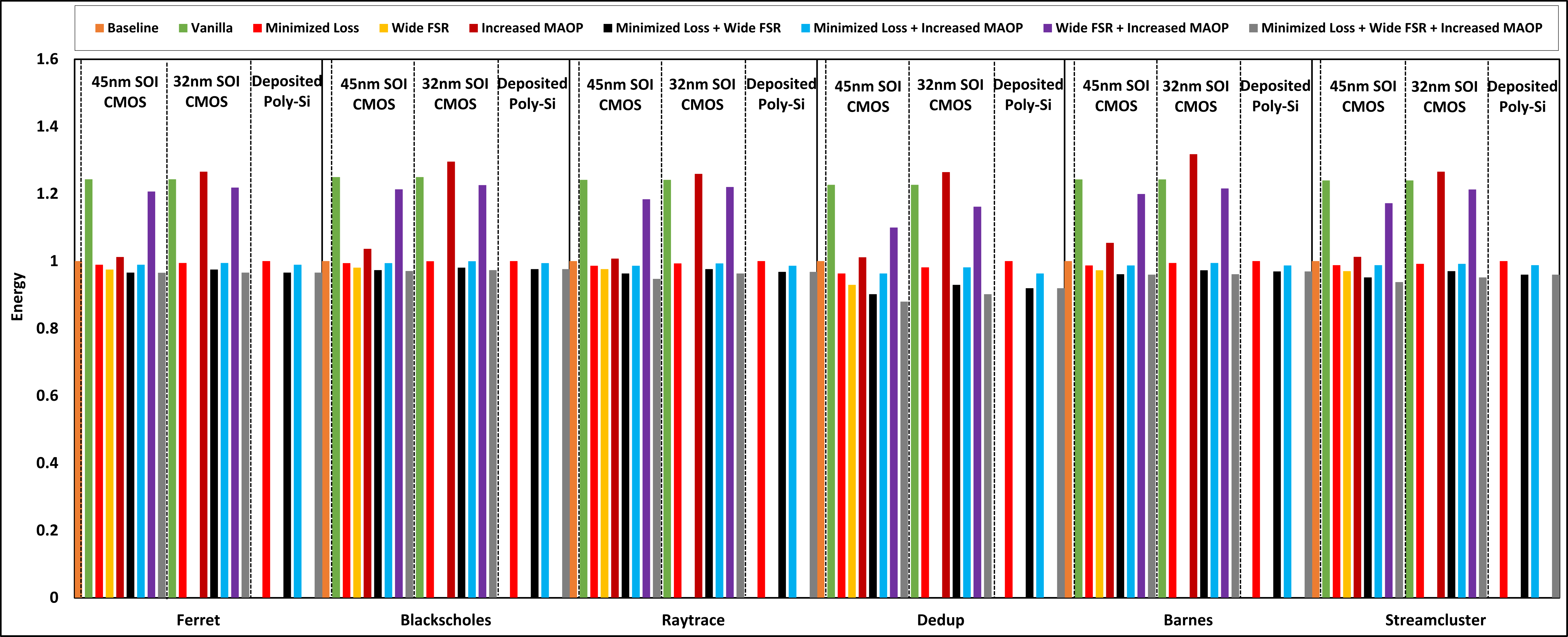}
    \caption{Energy comparison of on-SiPh variants derived from 45nm SOI CMOS, 32nm SOI CMOS and deposited poly-si photonic platforms implemented on NUPLet architecture.}
    \label{Fig:8}
\end{figure}

\begin{figure}
    \centering
    \includegraphics[scale = 0.168]{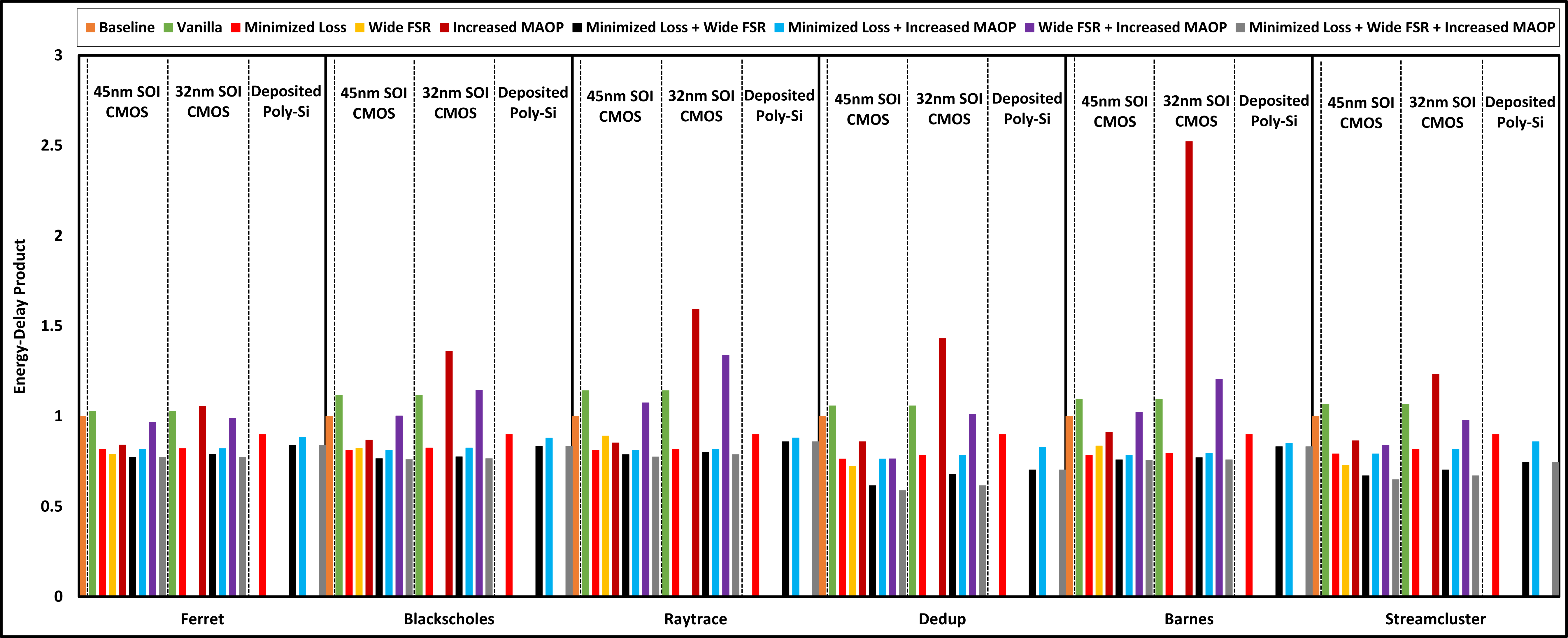}
    \caption{Energy-delay product comparison of on-SiPh variants derived from 45nm SOI CMOS, 32nm SOI CMOS and deposited poly-si photonic platforms implemented on NUPLet architecture.}
    \label{Fig:9}
\end{figure}

\begin{figure}[htp]

\subfloat[ResNet50]{%
  \includegraphics[scale = 0.25]{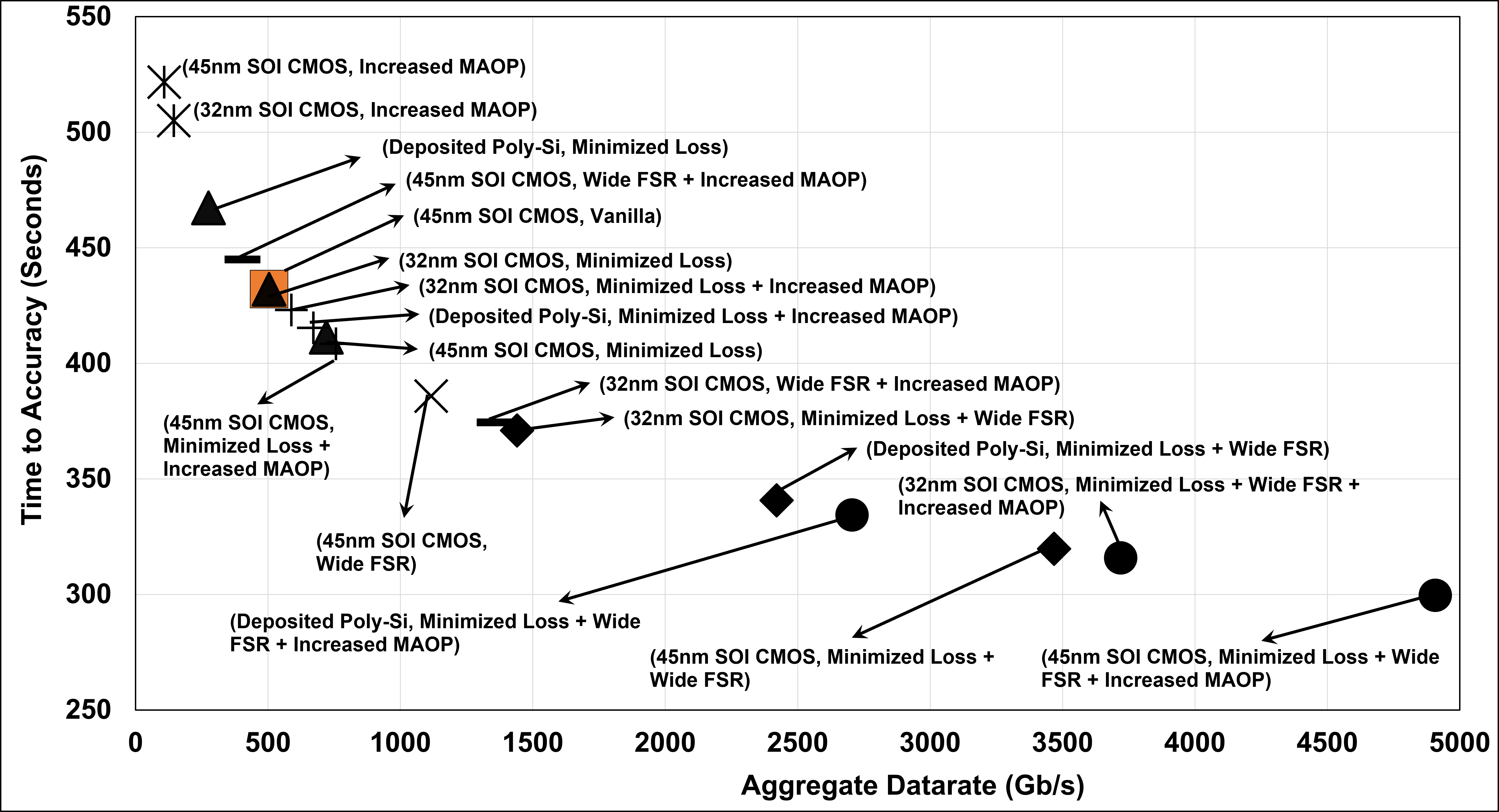}%
}

\subfloat[Transformer]{%
  \includegraphics[scale = 0.25]{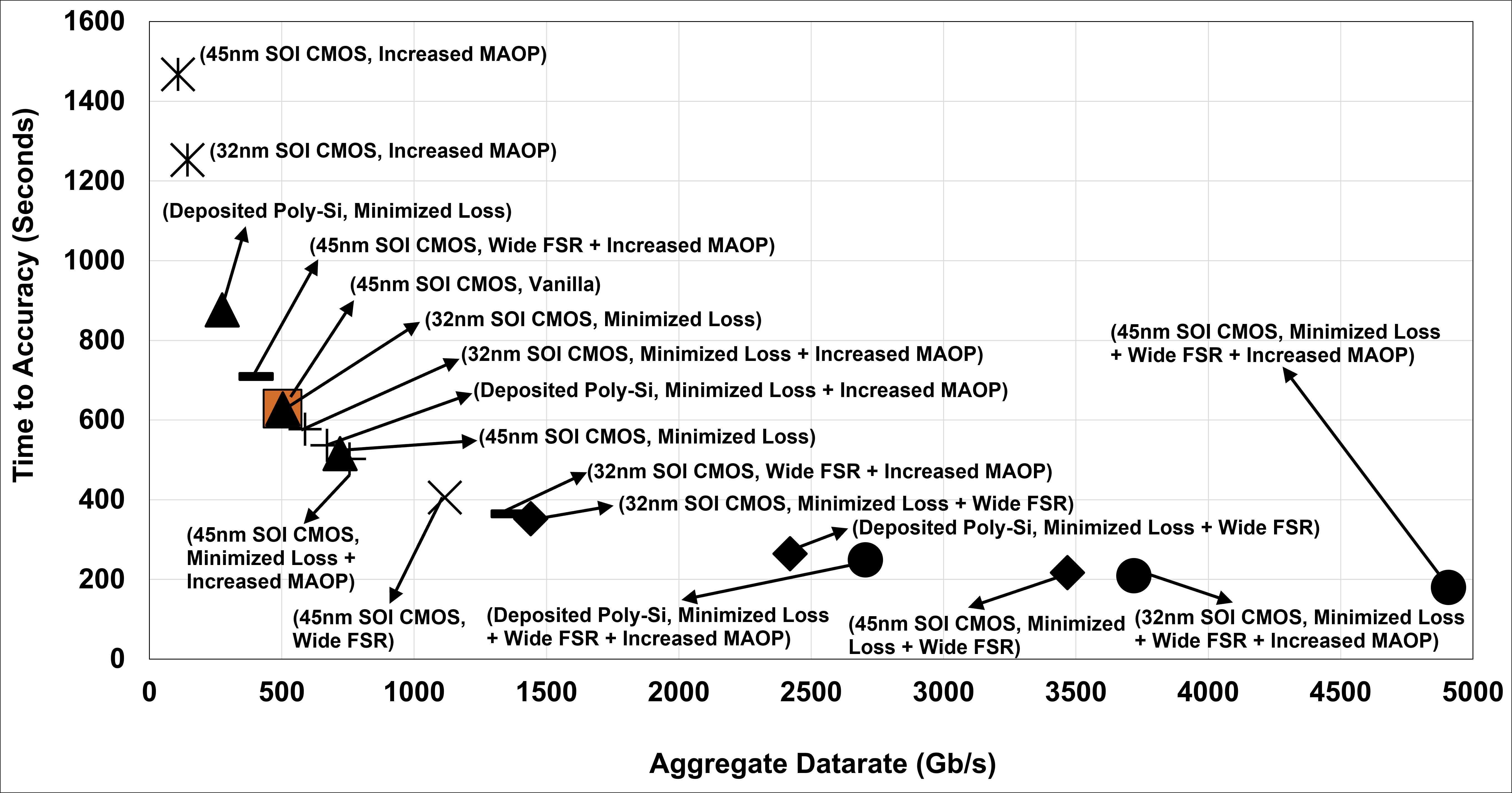}%
}

\subfloat[Megatron]{%
  \includegraphics[scale = 0.25]{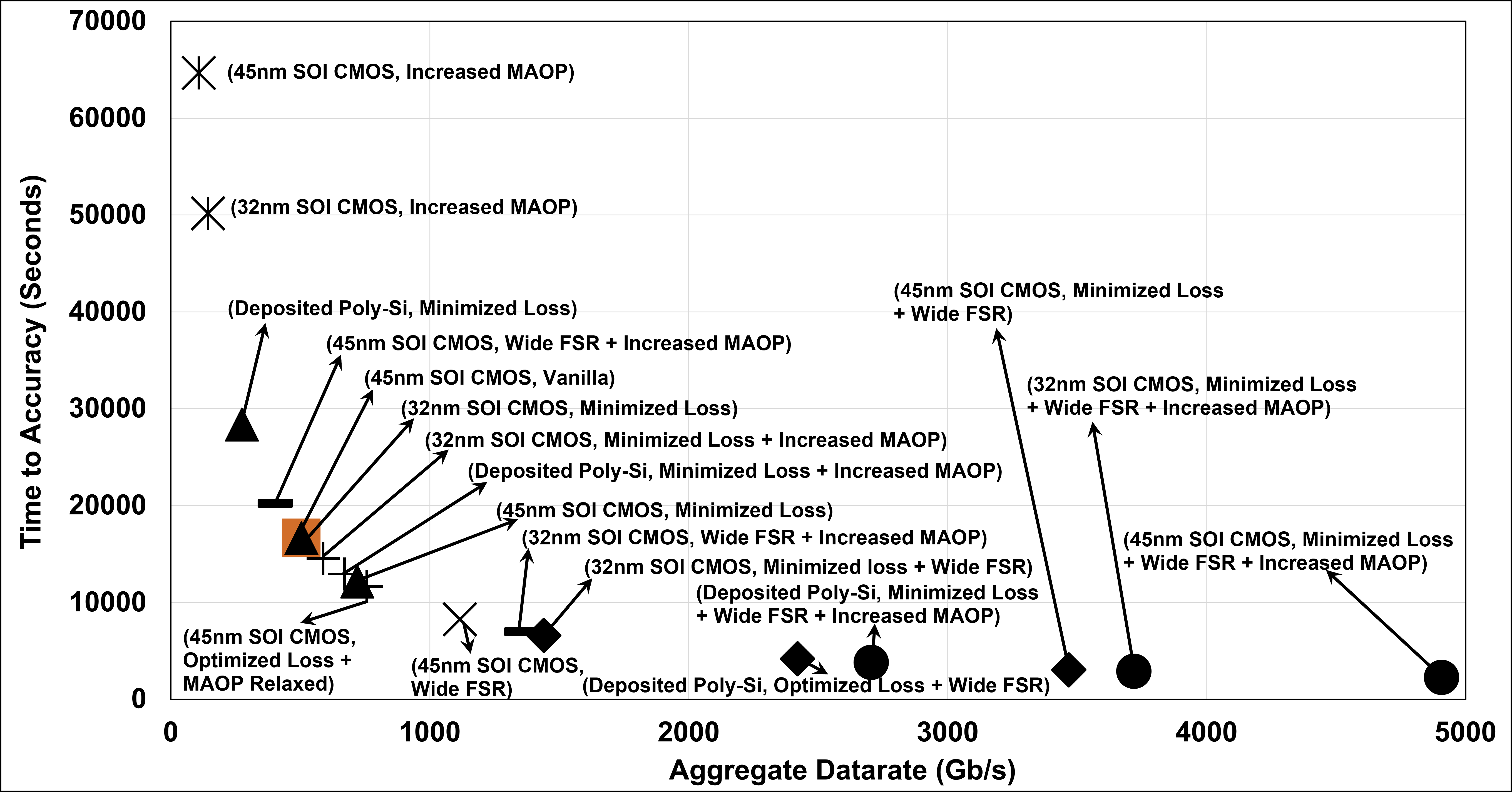}%
}

\caption{Impact of aggregate bandwidth on training time}
\label{Fig:10}
\end{figure}

\subsubsection{System-level Analysis on CPU based multi-core multi-chiplet module}
From the system-level analysis on NUPLet, we have evaluated performance, energy consumption and energy-delay product of NUPLet architecture employed with the derived on-SiPhI inter-chiplet variants. The longest inter-chiplet waveguide length we have considered for this analysis is 8 cm. For this waveguide length, \textit{Wide FSR} variant derived from 32nm SOI CMOS platform, and \textit{Vanilla}, \textit{Wide FSR}, \textit{Increased MAOP} and \textit{Wide FSR + Increased MAOP} variants corresponding to deposited poly-Si platform become non-viable due to high insertion loss (Fig. \ref{Fig:4} and Section 4.2.1). Performance, energy consumption and energy-delay product of the viable architecture variants are discussed below.

Fig. \ref{Fig:7}, Fig. \ref{Fig:8} and Fig. \ref{Fig:9} illustrate the relative performance (inverse of simulated execution time), energy consumption and energy-delay product of different variants of NUPLet architecture respectively corresponding to three different fabrication platforms for various PARSEC benchmark applications \cite{bienia2008}. The metric energy refers to the energy consumed by cores and lasers during the execution of an application. All the results are normalized to a baseline variant of NUPLet which has an N$_\lambda$ of 32 and bitrate of 10 Gb/s.  As we can infer from Fig. \ref{Fig:7}, among variants corresponding to 45nm SOI CMOS platform, the NUPLet architecture that employs \textit{Minimized Loss + Wide FSR + Increased MAOP}, \textit{Minimized Loss + Wide FSR}, \textit{Minimized Loss + Increased MAOP}, \textit{Minimized Loss} and \textit{Wide FSR} inter-chiplet variants achieve 33\%, 31.5\%, 23.6\%, 23.5\% and 22\% better performance on average respectively across all benchmark applications compared to the baseline variant. 

In terms of energy (Fig. \ref{Fig:8}), among variants corresponding to 45nm SOI CMOS platform, the NUPLet architecture that employs \textit{Minimized Loss + Wide FSR + Increased MAOP}, \textit{Minimized Loss + Wide FSR} variants consume 5.7\% and 5\% less energy on average respectively, followed by the NUPLet variants that employ \textit{Wide FSR}, \textit{Minimized Loss + Increased MAOP} and \textit{Minimized Loss} inter-chiplet variants, across all benchmark applications compared to the baseline variant. In terms of energy-delay product (Fig. \ref{Fig:9}), among inter-chiplet variants corresponding to 45nm SOI CMOS platform, the NUPLet architecture that employs \textit{Minimized Loss + Wide FSR + Increased MAOP}, \textit{Minimized Loss + Wide FSR} and \textit{Minimized Loss + Increased MAOP} inter-chiplet variants achieve 29\%, 27\% and 21\% less energy-delay product on average respectively followed by \textit{Wide FSR}, \textit{Minimized Loss} and \textit{Wide FSR} inter-chiplet variants across all benchmark applications compared to the baseline variant. 

Therefore, the NUPLet architecture that employs \textit{Minimized Loss + Wide FSR} and \textit{Minimized Loss + Wide FSR + Increased MAOP}  variants corresponding to 45nm SOI CMOS achieve better performance and incur less energy on average across all benchmark applications compared to the baseline variant. This is because of low insertion loss of inter-chiplet waveguides and high bandwidth of on-SiPhI inter-chiplet links, combined with NUCA and prediction schemes of NUPLet. This is leveraged by the ICOSs of the NUPLet to send more number of inter-chiplet messages/data packets at a time without any wastage of laser power, resulting in execution of application in less number of epochs with enhanced performance and less energy consumption.

Similarl, among the inter-chiplet variants corresponding to 32nm SOI CMOS platform, the NUPLet architecture that employs \textit{Minimized Loss + Wide FSR + Increased MAOP} and \textit{Minimized Loss + Wide FSR} inter-chiplet variants achieve 31.2\% and 28.6\% better performance respectively on average across all benchmark applications compared to baseline variant. This is followed by \textit{Minimized Loss + Increased MAOP} and \textit{Minimized Loss} inter-chiplet variants that achieve ~22\% better performance on average compared to the baseline variant. In terms of energy (Fig. \ref{Fig:8}), the NUPLet architecture that employs \textit{Minimized Loss + Wide FSR + Increased MAOP} and \textit{Minimized Loss + Wide FSR} inter-chiplet variants corresponding to 32nm SOI CMOS platform incur 5\% and 3.3\% less energy on average respectively across all benchmark applications compared to the baseline variant. In terms of energy-delay product (Fig. \ref{Fig:9}), \textit{Minimized Loss + Wide FSR + Increased MAOP} and \textit{Minimized Loss + Wide FSR} inter-chiplet variants corresponding to 32nm SOI CMOS platform achieve 27\% and 25\% less energy-delay product on average respectively across all benchmark applications compared to the baseline variant. Therefore, the NUPLet architecture that employs \textit{Minimized Loss + Wide FSR + Increased MAOP} and \textit{Minimized Loss + Wide FSR} variants corresponding to 32nm SOI CMOS platform achieve better performance and incur less energy on average across all benchmark applications compared to the baseline variant.

Similarly, among the inter-chiplet variants corresponding to deposited poly-Si platform, the NUPLet architecture that employs \textit{Minimized Loss + Wide FSR + Increased MAOP} and \textit{Minimized Loss + Wide FSR} inter-chiplet variants achieve 20\% better performance, consume 4\% less energy and achieve 20\% less energy-delay product on average respectively across all benchmark applications compared to the baseline variant. 

Therefore, from the system-level evaluation on NUPLet \cite{bashir2017}, we have observed that chiplet based PNoC architectures that employ \textit{Minimized Loss + Wide FSR + Increased MAOP}, \textit{Minimized Loss + Wide FSR} and \textit{Minimized Loss} on-SiPhI inter-chiplet variants corresponding to 45nm SOI CMOS, 32nm SOI CMOS and deposited poly-si platforms achieve superior performance and consume less energy compared to other inter-chiplet variants.

\subsubsection{System-level analysis on GPU based multi-chiplet module}
For the system-level analysis on GPU based MCM, we have utilized the simulator provided in \cite{khani2021} and evaluated time-to-accuracy i.e., the training time of three conventional DNN models namely ResNet50, Transformer and Megatron based on the aggregate bandwidth of our derived inter-chiplet variants enacted in GPU based MCMs. 

As we can infer from Fig. \ref{Fig:10}, GPU based MCMs that employ \textit{Minimized Loss + Wide FSR} and \textit{Minimized Loss + Wide FSR + Increased MAOP}  inter-chiplet variants corresponding to 45nm SOI CMOS, 32nm SOI CMOS and deposited poly-Si platforms enable at least 1-1.75$\times$, 2-8$\times$, 4-30$\times$ faster training time for ResNet50, Transformer and Megatron respectively. This is because, both of these inter-chiplet variants achieve multi-Tb/s aggregate bandwidth at link-level.

\section{Summary}
In this section, we summarize the results obtained from the link-level and system-level analysis of on-SiPhI inter-chiplet variants derived based on our identified design pathways (Table \ref{Table:1}), corresponding to three different SiPh fabrication platforms (Table \ref{Table:2}).

\subsection{Link-Level Evaluation}
From the link-level analysis, we have evaluated the aggregate bandwidth (primary Y-axis in Fig. \ref{Fig:4}) and EPB (secondary Y-axis in Fig. \ref{Fig:4}) for different on-SiPhI inter-chiplet variants corresponding to three different SiPh fabrication platforms, for different waveguide lengths (X-axis in Fig. \ref{Fig:4}). Based on the results obtained from link-level analysis, we have classified the derived inter-chiplet variants into two categories namely non-viable variants and viable variants.

\subsubsection{Non-Viable Variants}
Non-viable variants are the inter-chiplet variants that do not support any wavelength channels in the link and therefore support no aggregate bandwidth for longer waveguide lengths. The non-viable variants determined from this analysis are as follows:

\begin{enumerate}
    \item \textit{Vanilla} and \textit{Wide FSR} variants corresponding to 32nm SOI CMOS and deposited poly-Si SiPh platforms do not support any wavelength channels due to high insertion loss in the link whereas the same variants corresponding to 45nm SOI CMOS platform support wavelength channels up to a link length of 4cm and they can be made viable for longer waveguide lengths by employing repeaters
    \item \textit{Increased MAOP} and \textit{Wide FSR + Increased MAOP} variants corresponding to deposited poly-Si platform does not support any wavelength channels due to high insertion loss in the link whereas the same variants corresponding to 45nm SOI CMOS and 32nm SOI CMOS platforms supports wavelength channels up to a link length of 8 cm and 2 cm respectively and they can be made viable by utilizing repeaters
\end{enumerate}

\subsubsection{Viable Variants}
Viable variants are the inter-chiplet variants that support wavelength channels in the link up to link lengths as long as 10cm. The viable variants determined from this analysis are as follows:

\begin{enumerate}
    \item \textit{Minimized Loss}, \textit{Minimized Loss + Wide FSR}, \textit{Minimized Loss + Increased MAOP} and \textit{Minimized Loss + Wide FSR + Increased MAOP} variants corresponding to three different SiPh fabrication platforms support wavelength channels up to a link length of 10cm
    \item Among the viable variants, \textit{Minimized Loss + Wide FSR + Increased MAOP} variant corresponding to 45nm SOI CMOS platform achieves highest aggregate bandwidth of 4.92 Tb/s with corresponding EPB of 9.2pJ/bit whereas \textit{Minimized Loss + Wide FSR} variant corresponding to the same fabrication platform achieves lowest EPB of 0.218 pJ/bit with corresponding aggregate bandwidth of 4.6 Tb/s 
\end{enumerate}

\subsection{System-Level Evaluation}
We have implemented the on-SiPhI inter-chiplet variants on a CPU based multi-core multi-chiplet architecture named NUPLet \cite{bashir2017} and a GPU based multi-chiplet module (MCM) \cite{khani2021} and performed a system-level analysis. Results of this analysis are summarized as follows.

\subsubsection{System-Level Evaluation on CPU Based Multi-Core Multi-Chiplet Architecture}
We have implemented the derived inter-chiplet variants on NUPLet architecture\cite{bashir2017} and performed a benchmark-driven simulation based analysis from which we have evaluated the performance (Fig.; \ref{Fig:7}), energy consumption (Fig. \ref{Fig:8}) and energy-delay product (Fig. \ref{Fig:9}) of the NUPLet architecture. The results obtained from this evaluation are summarized as follows:

\begin{enumerate}
    \item NUPLet architecture that employs \textit{Minimized Loss}, \textit{Minimized Loss + Wide FSR} and \textit{Minimized Loss + Wide FSR + Increased MAOP} inter-chiplet variants corresponding to three considered SiPh fabrication platforms (Table \ref{Table:2}) achieve superior performance and consume less energy compared to other inter-chiplet variants
\end{enumerate}

\subsubsection{System-Level Evaluation on GPU Based Multi-Chiplet Module}
We have implemented the derived inter-chiplet variants on a GPU based MCM \cite{khani2021} and performed a system-level analysis utilizing the simulator provided in \cite{khani2021}, from which we have evaluated the time-to-accuracy of three conventional DNN models namely ResNet50 (Fig. \ref{Fig:10}(a)), Transformer (Fig. \ref{Fig:10}(b)) and Megatron (Fig. \ref{Fig:10}(c)). The results of this evaluation are summarized as follows:

\begin{enumerate}
    \item GPU based MCMs that employ \textit{Minimized Loss + Wide FSR} and \textit{Minimized Loss + Wide FSR + Increased MAOP} inter-chiplet variants corresponding to the three considered SiPh fabrication platforms (Table \ref{Table:2}) accelerate the training time for ResNet50, Transformer and Megatron DNN models by at least 1-1.75$\times$, 2-8$\times$ and 4-30$\times$ respectively.
\end{enumerate}

\section{Summary and Conclusion}
The dwindling of Moore’s law has drastically increased the complexity and the cost of fabricating large-scale, monolithic Systems-on-Chip (SoCs). Therefore, the industry has adopted fragmentation of monolithic SoCs into several smaller chiplets, which are then assembled using silicon interposer. However, to meet the growing demands of modern data-centric workloads, it is vital to realize on-interposer inter-chiplet communication bandwidth of multi-Tb/s and end-to-end communication latency of <10ns. To meet these bandwidth and latency goals, prior works have focused on a potential solution of using the silicon photonic interposer (SiPhI) for integrating and interconnecting a large number of chiplets into a system-in-package (SiP). However, the designs of on-SiPhI interconnects, demonstrated so far, have to still evolve swiftly in order to meet the goal of multi-Tb/s bandwidth. But the possible design pathways that can aid in such evolution, have not been explored yet. Therefore, in this paper, we identified several design pathways that can aid on-SiPhI inter-chiplet interconnects to meet the goal of achieving multi-Tb/s bandwidth.

Based on the identified design pathways and three different photonic fabrication platforms, namely 45nm SOI CMOS, 32nm SOI CMOS and deposited poly-Si, we derived twenty four design variants of on-SiPhI inter-chiplet interconnects. Then, we performed an extensive link-level and system-level analysis for each of these variants. From the link-level analysis, we observed that the design pathways that simultaneously enhance the spectral range and optical power budget available for wavelength division multiplexing provide enough impetus to the corresponding on-SiPhI inter-chiplet links to achieve aggregate bandwidth of >4Tb/s. Based on the link-level analysis, we performed system-level analysis from which we observed that the state-of-the-art CPU and GPU based SiPs that employ such multi-Tb/s on-SiPhI inter-chiplet links achieve significantly improved performance and energy-efficiency. 


\bibliographystyle{ACM-Reference-Format}
\bibliography{ref}


\begin{thebibliography}{88}


\ifx \showCODEN    \undefined \def \showCODEN     #1{\unskip}     \fi
\ifx \showDOI      \undefined \def \showDOI       #1{#1}\fi
\ifx \showISBNx    \undefined \def \showISBNx     #1{\unskip}     \fi
\ifx \showISBNxiii \undefined \def \showISBNxiii  #1{\unskip}     \fi
\ifx \showISSN     \undefined \def \showISSN      #1{\unskip}     \fi
\ifx \showLCCN     \undefined \def \showLCCN      #1{\unskip}     \fi
\ifx \shownote     \undefined \def \shownote      #1{#1}          \fi
\ifx \showarticletitle \undefined \def \showarticletitle #1{#1}   \fi
\ifx \showURL      \undefined \def \showURL       {\relax}        \fi
\providecommand\bibfield[2]{#2}
\providecommand\bibinfo[2]{#2}
\providecommand\natexlab[1]{#1}
\providecommand\showeprint[2][]{arXiv:#2}

\bibitem[Abrams et~al\mbox{.}(2020)]%
        {abrams2020}
\bibfield{author}{\bibinfo{person}{Nathan~C Abrams}, \bibinfo{person}{Qixiang
  Cheng}, \bibinfo{person}{Madeleine Glick}, \bibinfo{person}{Moises Jezzini},
  \bibinfo{person}{Padraic Morrissey}, \bibinfo{person}{Peter O'Brien}, {and}
  \bibinfo{person}{Keren Bergman}.} \bibinfo{year}{2020}\natexlab{}.
\newblock \showarticletitle{Silicon photonic 2.5 D multi-chip module
  transceiver for high-performance data centers}.
\newblock \bibinfo{journal}{\emph{Journal of Lightwave Technology}}
  \bibinfo{volume}{38}, \bibinfo{number}{13} (\bibinfo{year}{2020}),
  \bibinfo{pages}{3346--3357}.
\newblock


\bibitem[Arunkumar et~al\mbox{.}(2017)]%
        {arunkumar2017}
\bibfield{author}{\bibinfo{person}{Akhil Arunkumar}, \bibinfo{person}{Evgeny
  Bolotin}, \bibinfo{person}{Benjamin Cho}, \bibinfo{person}{Ugljesa Milic},
  \bibinfo{person}{Eiman Ebrahimi}, \bibinfo{person}{Oreste Villa},
  \bibinfo{person}{Aamer Jaleel}, \bibinfo{person}{Carole-Jean Wu}, {and}
  \bibinfo{person}{David Nellans}.} \bibinfo{year}{2017}\natexlab{}.
\newblock \showarticletitle{MCM-GPU: Multi-chip-module GPUs for continued
  performance scalability}.
\newblock \bibinfo{journal}{\emph{ACM SIGARCH Computer Architecture News}}
  \bibinfo{volume}{45}, \bibinfo{number}{2} (\bibinfo{year}{2017}),
  \bibinfo{pages}{320--332}.
\newblock


\bibitem[Atabaki et~al\mbox{.}(2018)]%
        {atabaki2018}
\bibfield{author}{\bibinfo{person}{Amir~H Atabaki}, \bibinfo{person}{Sajjad
  Moazeni}, \bibinfo{person}{Fabio Pavanello}, \bibinfo{person}{Hayk
  Gevorgyan}, \bibinfo{person}{Jelena Notaros}, \bibinfo{person}{Luca
  Alloatti}, \bibinfo{person}{Mark~T Wade}, \bibinfo{person}{Chen Sun},
  \bibinfo{person}{Seth~A Kruger}, \bibinfo{person}{Huaiyu Meng},
  {et~al\mbox{.}}} \bibinfo{year}{2018}\natexlab{}.
\newblock \showarticletitle{Integrating photonics with silicon nanoelectronics
  for the next generation of systems on a chip}.
\newblock \bibinfo{journal}{\emph{Nature}} \bibinfo{volume}{556},
  \bibinfo{number}{7701} (\bibinfo{year}{2018}), \bibinfo{pages}{349--354}.
\newblock


\bibitem[Bahadori and Bergman(2018)]%
        {bahadori2018}
\bibfield{author}{\bibinfo{person}{Meisam Bahadori} {and}
  \bibinfo{person}{Keren Bergman}.} \bibinfo{year}{2018}\natexlab{}.
\newblock \showarticletitle{Low-power optical interconnects based on resonant
  silicon photonic devices: Recent advances and challenges}. In
  \bibinfo{booktitle}{\emph{Proceedings of the 2018 on Great Lakes Symposium on
  VLSI}}. \bibinfo{pages}{305--310}.
\newblock


\bibitem[Bahadori et~al\mbox{.}(2016a)]%
        {bahadori2016}
\bibfield{author}{\bibinfo{person}{Meisam Bahadori},
  \bibinfo{person}{S{\'e}bastien Rumley}, \bibinfo{person}{Hasitha
  Jayatilleka}, \bibinfo{person}{Kyle Murray}, \bibinfo{person}{Nicolas~AF
  Jaeger}, \bibinfo{person}{Lukas Chrostowski}, \bibinfo{person}{Sudip
  Shekhar}, {and} \bibinfo{person}{Keren Bergman}.}
  \bibinfo{year}{2016}\natexlab{a}.
\newblock \showarticletitle{Crosstalk penalty in microring-based silicon
  photonic interconnect systems}.
\newblock \bibinfo{journal}{\emph{Journal of Lightwave Technology}}
  \bibinfo{volume}{34}, \bibinfo{number}{17} (\bibinfo{year}{2016}),
  \bibinfo{pages}{4043--4052}.
\newblock


\bibitem[Bahadori et~al\mbox{.}(2016b)]%
        {bahadori2016design}
\bibfield{author}{\bibinfo{person}{Meisam Bahadori},
  \bibinfo{person}{S{\'e}bastien Rumley}, \bibinfo{person}{Dessislava
  Nikolova}, {and} \bibinfo{person}{Keren Bergman}.}
  \bibinfo{year}{2016}\natexlab{b}.
\newblock \showarticletitle{Comprehensive design space exploration of silicon
  photonic interconnects}.
\newblock \bibinfo{journal}{\emph{Journal of Lightwave Technology}}
  \bibinfo{volume}{34}, \bibinfo{number}{12} (\bibinfo{year}{2016}),
  \bibinfo{pages}{2975--2987}.
\newblock


\bibitem[Bahadori et~al\mbox{.}(2017)]%
        {bahadori2017}
\bibfield{author}{\bibinfo{person}{Meisam Bahadori},
  \bibinfo{person}{S{\'e}bastien Rumley}, \bibinfo{person}{Robert Polster},
  \bibinfo{person}{Alexander Gazman}, \bibinfo{person}{Matt Traverso},
  \bibinfo{person}{Mark Webster}, \bibinfo{person}{Kaushik Patel}, {and}
  \bibinfo{person}{Keren Bergman}.} \bibinfo{year}{2017}\natexlab{}.
\newblock \showarticletitle{Energy-performance optimized design of silicon
  photonic interconnection networks for high-performance computing}. In
  \bibinfo{booktitle}{\emph{Design, Automation \& Test in Europe Conference \&
  Exhibition (DATE), 2017}}. IEEE, \bibinfo{pages}{326--331}.
\newblock


\bibitem[Bajwa et~al\mbox{.}(2017)]%
        {bajwa2017}
\bibfield{author}{\bibinfo{person}{Adeel~A Bajwa}, \bibinfo{person}{SivaChandra
  Jangam}, \bibinfo{person}{Saptadeep Pal}, \bibinfo{person}{Niteesh Marathe},
  \bibinfo{person}{Tingyu Bai}, \bibinfo{person}{Takafumi Fukushima},
  \bibinfo{person}{Mark Goorsky}, {and} \bibinfo{person}{Subramanian~S Iyer}.}
  \bibinfo{year}{2017}\natexlab{}.
\newblock \showarticletitle{Heterogeneous integration at fine pitch (≤ 10
  $\mu$m) using thermal compression bonding}. In \bibinfo{booktitle}{\emph{2017
  IEEE 67th electronic components and technology conference (ECTC)}}. IEEE,
  \bibinfo{pages}{1276--1284}.
\newblock


\bibitem[Basak et~al\mbox{.}(2008)]%
        {basak2008high}
\bibfield{author}{\bibinfo{person}{J Basak}, \bibinfo{person}{L Liao},
  \bibinfo{person}{A Liu}, \bibinfo{person}{H Nguyen}, \bibinfo{person}{M
  Paniccia}, \bibinfo{person}{Yoel Chetrit}, {and} \bibinfo{person}{Doron
  Rubin}.} \bibinfo{year}{2008}\natexlab{}.
\newblock \showarticletitle{High speed photonics on an SOI platform}. In
  \bibinfo{booktitle}{\emph{2008 IEEE International SOI Conference}}. IEEE,
  \bibinfo{pages}{85--86}.
\newblock


\bibitem[Bashir et~al\mbox{.}(2019)]%
        {bashir2019}
\bibfield{author}{\bibinfo{person}{Janibul Bashir}, \bibinfo{person}{Eldhose
  Peter}, {and} \bibinfo{person}{Smruti~R Sarangi}.}
  \bibinfo{year}{2019}\natexlab{}.
\newblock \showarticletitle{A survey of on-chip optical interconnects}.
\newblock \bibinfo{journal}{\emph{ACM Computing Surveys (CSUR)}}
  \bibinfo{volume}{51}, \bibinfo{number}{6} (\bibinfo{year}{2019}),
  \bibinfo{pages}{1--34}.
\newblock


\bibitem[Bashir and Sarangi(2017)]%
        {bashir2017}
\bibfield{author}{\bibinfo{person}{Janibul Bashir} {and}
  \bibinfo{person}{Smruti~R Sarangi}.} \bibinfo{year}{2017}\natexlab{}.
\newblock \showarticletitle{NUPLet: A photonic based multi-chip NUCA
  architecture}. In \bibinfo{booktitle}{\emph{2017 IEEE International
  Conference on Computer Design (ICCD)}}. IEEE, \bibinfo{pages}{617--624}.
\newblock


\bibitem[Bergman et~al\mbox{.}(2014)]%
        {bergman2014}
\bibfield{author}{\bibinfo{person}{Keren Bergman}, \bibinfo{person}{Luca~P
  Carloni}, \bibinfo{person}{Aleksandr Biberman}, \bibinfo{person}{Johnnie
  Chan}, {and} \bibinfo{person}{Gilbert Hendry}.}
  \bibinfo{year}{2014}\natexlab{}.
\newblock \bibinfo{booktitle}{\emph{Photonic network-on-chip design}}.
\newblock \bibinfo{publisher}{Springer}.
\newblock


\bibitem[Bharadwaj et~al\mbox{.}(2020)]%
        {bharadwaj2020}
\bibfield{author}{\bibinfo{person}{Srikant Bharadwaj}, \bibinfo{person}{Jieming
  Yin}, \bibinfo{person}{Bradford Beckmann}, {and} \bibinfo{person}{Tushar
  Krishna}.} \bibinfo{year}{2020}\natexlab{}.
\newblock \showarticletitle{Kite: A family of heterogeneous interposer
  topologies enabled via accurate interconnect modeling}. In
  \bibinfo{booktitle}{\emph{2020 57th ACM/IEEE Design Automation Conference
  (DAC)}}. IEEE, \bibinfo{pages}{1--6}.
\newblock


\bibitem[Bienia et~al\mbox{.}(2008)]%
        {bienia2008}
\bibfield{author}{\bibinfo{person}{Christian Bienia}, \bibinfo{person}{Sanjeev
  Kumar}, \bibinfo{person}{Jaswinder~Pal Singh}, {and} \bibinfo{person}{Kai
  Li}.} \bibinfo{year}{2008}\natexlab{}.
\newblock \showarticletitle{The PARSEC benchmark suite: Characterization and
  architectural implications}. In \bibinfo{booktitle}{\emph{Proceedings of the
  17th international conference on Parallel architectures and compilation
  techniques}}. \bibinfo{pages}{72--81}.
\newblock


\bibitem[Borghi et~al\mbox{.}(2021)]%
        {borghi2021}
\bibfield{author}{\bibinfo{person}{Massimo Borghi}, \bibinfo{person}{Davide
  Bazzanella}, \bibinfo{person}{Mattia Mancinelli}, {and}
  \bibinfo{person}{Lorenzo Pavesi}.} \bibinfo{year}{2021}\natexlab{}.
\newblock \showarticletitle{On the modeling of thermal and free carrier
  nonlinearities in silicon-on-insulator microring resonators}.
\newblock \bibinfo{journal}{\emph{Optics Express}} \bibinfo{volume}{29},
  \bibinfo{number}{3} (\bibinfo{year}{2021}), \bibinfo{pages}{4363--4377}.
\newblock


\bibitem[Chen et~al\mbox{.}(2014)]%
        {chen2014}
\bibfield{author}{\bibinfo{person}{Yunji Chen}, \bibinfo{person}{Tao Luo},
  \bibinfo{person}{Shaoli Liu}, \bibinfo{person}{Shijin Zhang},
  \bibinfo{person}{Liqiang He}, \bibinfo{person}{Jia Wang},
  \bibinfo{person}{Ling Li}, \bibinfo{person}{Tianshi Chen},
  \bibinfo{person}{Zhiwei Xu}, \bibinfo{person}{Ninghui Sun}, {et~al\mbox{.}}}
  \bibinfo{year}{2014}\natexlab{}.
\newblock \showarticletitle{Dadiannao: A machine-learning supercomputer}. In
  \bibinfo{booktitle}{\emph{2014 47th Annual IEEE/ACM International Symposium
  on Microarchitecture}}. IEEE, \bibinfo{pages}{609--622}.
\newblock


\bibitem[Chuang et~al\mbox{.}(2013)]%
        {chuang2013}
\bibfield{author}{\bibinfo{person}{Yi-Lin Chuang}, \bibinfo{person}{Chung-Sheng
  Yuan}, \bibinfo{person}{Ji-Jan Chen}, \bibinfo{person}{Ching-Fang Chen},
  \bibinfo{person}{Ching-Shun Yang}, \bibinfo{person}{Wei-Pin Changchien},
  \bibinfo{person}{Charles~CC Liu}, {and} \bibinfo{person}{Frank Lee}.}
  \bibinfo{year}{2013}\natexlab{}.
\newblock \showarticletitle{Unified methodology for heterogeneous integration
  with CoWoS technology}. In \bibinfo{booktitle}{\emph{2013 IEEE 63rd
  Electronic Components and Technology Conference}}. IEEE,
  \bibinfo{pages}{852--859}.
\newblock


\bibitem[Chung et~al\mbox{.}(2017)]%
        {chung2017}
\bibfield{author}{\bibinfo{person}{Eric Chung}, \bibinfo{person}{Jeremy
  Fowers}, \bibinfo{person}{Kalin Ovtcharov}, \bibinfo{person}{Michael
  Papamichael}, \bibinfo{person}{Adrian Caulfield}, \bibinfo{person}{Todd
  Massengil}, \bibinfo{person}{Ming Liu}, \bibinfo{person}{Daniel Lo},
  \bibinfo{person}{Shlomi Alkalay}, \bibinfo{person}{Michael Haselman},
  {et~al\mbox{.}}} \bibinfo{year}{2017}\natexlab{}.
\newblock \showarticletitle{Accelerating persistent neural networks at
  datacenter scale}. In \bibinfo{booktitle}{\emph{Hot Chips}},
  Vol.~\bibinfo{volume}{29}.
\newblock


\bibitem[Corporation(2020a)]%
        {Intel}
\bibfield{author}{\bibinfo{person}{Intel Corporation}.}
  \bibinfo{year}{2020}\natexlab{a}.
\newblock \bibinfo{booktitle}{\emph{Architecture Day 2020}}.
\newblock
\urldef\tempurl%
\url{https://newsroom.intel.com/wp-content/uploads/sites/11/2020/08/Intel-Architecture-Day-2020-Presentation-Slides.pdf}
\showURL{%
\tempurl}


\bibitem[Corporation(2020b)]%
        {semiconductor}
\bibfield{author}{\bibinfo{person}{Semiconductor~Research Corporation}.}
  \bibinfo{year}{2020}\natexlab{b}.
\newblock \bibinfo{booktitle}{\emph{The Decadal Plan for Semiconductors}}.
\newblock
\urldef\tempurl%
\url{https://www.src.org/about/decadal-plan/}
\showURL{%
Retrieved March 23, 2022 from \tempurl}


\bibitem[DARPA(2018)]%
        {DARPA}
\bibfield{author}{\bibinfo{person}{DARPA}.} \bibinfo{year}{2018}\natexlab{}.
\newblock \bibinfo{booktitle}{\emph{PIPES}}.
\newblock
\urldef\tempurl%
\url{https://s3-us-west-2.amazonaws.com/instrumentl/grantsgov/310031.pdf}
\showURL{%
\tempurl}


\bibitem[Daudlin et~al\mbox{.}(2021)]%
        {daudlin2021}
\bibfield{author}{\bibinfo{person}{Stuart Daudlin}, \bibinfo{person}{Anthony
  Rizzo}, \bibinfo{person}{Nathan~C Abrams}, \bibinfo{person}{Sunwoo Lee},
  \bibinfo{person}{Devesh Khilwani}, \bibinfo{person}{Vaishnavi Murthy},
  \bibinfo{person}{James Robinson}, \bibinfo{person}{Terence Collier},
  \bibinfo{person}{Alyosha Molnar}, {and} \bibinfo{person}{Keren Bergman}.}
  \bibinfo{year}{2021}\natexlab{}.
\newblock \showarticletitle{3D-Integrated Multichip Module Transceiver for
  Terabit-Scale DWDM Interconnects}. In \bibinfo{booktitle}{\emph{Optical Fiber
  Communication Conference}}. Optical Society of America,
  \bibinfo{pages}{Th4A--4}.
\newblock


\bibitem[De~Cea et~al\mbox{.}(2019)]%
        {de2019}
\bibfield{author}{\bibinfo{person}{Marc De~Cea}, \bibinfo{person}{Amir~H
  Atabaki}, {and} \bibinfo{person}{Rajeev~J Ram}.}
  \bibinfo{year}{2019}\natexlab{}.
\newblock \showarticletitle{Power handling of silicon microring modulators}.
\newblock \bibinfo{journal}{\emph{Optics express}} \bibinfo{volume}{27},
  \bibinfo{number}{17} (\bibinfo{year}{2019}), \bibinfo{pages}{24274--24285}.
\newblock


\bibitem[Dong et~al\mbox{.}(2017)]%
        {dong2017}
\bibfield{author}{\bibinfo{person}{Bowei Dong}, \bibinfo{person}{Xin Guo},
  \bibinfo{person}{Chong~Pei Ho}, \bibinfo{person}{Bo Li},
  \bibinfo{person}{Hong Wang}, \bibinfo{person}{Chengkuo Lee},
  \bibinfo{person}{Xianshu Luo}, {and} \bibinfo{person}{Guo-Qiang Lo}.}
  \bibinfo{year}{2017}\natexlab{}.
\newblock \showarticletitle{Silicon-on-insulator waveguide devices for
  broadband mid-infrared photonics}.
\newblock \bibinfo{journal}{\emph{IEEE Photonics Journal}} \bibinfo{volume}{9},
  \bibinfo{number}{3} (\bibinfo{year}{2017}), \bibinfo{pages}{1--10}.
\newblock


\bibitem[Dong et~al\mbox{.}(2010)]%
        {dong}
\bibfield{author}{\bibinfo{person}{Po Dong}, \bibinfo{person}{Wei Qian},
  \bibinfo{person}{Shirong Liao}, \bibinfo{person}{Hong Liang},
  \bibinfo{person}{Cheng-Chih Kung}, \bibinfo{person}{Ning-Ning Feng},
  \bibinfo{person}{Roshanak Shafiiha}, \bibinfo{person}{Joan Fong},
  \bibinfo{person}{Dazeng Feng}, \bibinfo{person}{Ashok~V Krishnamoorthy},
  {et~al\mbox{.}}} \bibinfo{year}{2010}\natexlab{}.
\newblock \showarticletitle{Low loss shallow-ridge silicon waveguides}.
\newblock \bibinfo{journal}{\emph{Optics express}} \bibinfo{volume}{18},
  \bibinfo{number}{14} (\bibinfo{year}{2010}), \bibinfo{pages}{14474--14479}.
\newblock


\bibitem[Eid et~al\mbox{.}(2016)]%
        {eid2016}
\bibfield{author}{\bibinfo{person}{Nourhan Eid}, \bibinfo{person}{Robert
  Boeck}, \bibinfo{person}{Hasitha Jayatilleka}, \bibinfo{person}{Lukas
  Chrostowski}, \bibinfo{person}{Wei Shi}, {and} \bibinfo{person}{Nicolas~AF
  Jaeger}.} \bibinfo{year}{2016}\natexlab{}.
\newblock \showarticletitle{FSR-free silicon-on-insulator microring resonator
  based filter with bent contra-directional couplers}.
\newblock \bibinfo{journal}{\emph{Optics express}} \bibinfo{volume}{24},
  \bibinfo{number}{25} (\bibinfo{year}{2016}), \bibinfo{pages}{29009--29021}.
\newblock


\bibitem[Fowers et~al\mbox{.}(2018)]%
        {fowers2018}
\bibfield{author}{\bibinfo{person}{Jeremy Fowers}, \bibinfo{person}{Kalin
  Ovtcharov}, \bibinfo{person}{Michael Papamichael}, \bibinfo{person}{Todd
  Massengill}, \bibinfo{person}{Ming Liu}, \bibinfo{person}{Daniel Lo},
  \bibinfo{person}{Shlomi Alkalay}, \bibinfo{person}{Michael Haselman},
  \bibinfo{person}{Logan Adams}, \bibinfo{person}{Mahdi Ghandi},
  {et~al\mbox{.}}} \bibinfo{year}{2018}\natexlab{}.
\newblock \showarticletitle{A configurable cloud-scale DNN processor for
  real-time AI}. In \bibinfo{booktitle}{\emph{2018 ACM/IEEE 45th Annual
  International Symposium on Computer Architecture (ISCA)}}. IEEE,
  \bibinfo{pages}{1--14}.
\newblock


\bibitem[Gaeta et~al\mbox{.}(2019)]%
        {gaeta2019}
\bibfield{author}{\bibinfo{person}{Alexander~L Gaeta}, \bibinfo{person}{Michal
  Lipson}, {and} \bibinfo{person}{Tobias~J Kippenberg}.}
  \bibinfo{year}{2019}\natexlab{}.
\newblock \showarticletitle{Photonic-chip-based frequency combs}.
\newblock \bibinfo{journal}{\emph{nature photonics}} \bibinfo{volume}{13},
  \bibinfo{number}{3} (\bibinfo{year}{2019}), \bibinfo{pages}{158--169}.
\newblock


\bibitem[Griffel(2000)]%
        {griffel2000}
\bibfield{author}{\bibinfo{person}{Giora Griffel}.}
  \bibinfo{year}{2000}\natexlab{}.
\newblock \showarticletitle{Vernier effect in asymmetrical ring resonator
  arrays}.
\newblock \bibinfo{journal}{\emph{IEEE Photonics Technology Letters}}
  \bibinfo{volume}{12}, \bibinfo{number}{12} (\bibinfo{year}{2000}),
  \bibinfo{pages}{1642--1644}.
\newblock


\bibitem[Gwennap(2018)]%
        {gwennap2018}
\bibfield{author}{\bibinfo{person}{Linley Gwennap}.}
  \bibinfo{year}{2018}\natexlab{}.
\newblock \showarticletitle{Graphcore makes big AI splash}.
\newblock \bibinfo{journal}{\emph{Microprocessor Rep., The Linley Group,
  Mountain View, CA, USA}} (\bibinfo{year}{2018}).
\newblock


\bibitem[He et~al\mbox{.}(2021)]%
        {he2021}
\bibfield{author}{\bibinfo{person}{An He}, \bibinfo{person}{Xuhan Guo},
  \bibinfo{person}{Ting Wang}, {and} \bibinfo{person}{Yikai Su}.}
  \bibinfo{year}{2021}\natexlab{}.
\newblock \showarticletitle{Ultracompact Fiber-to-Chip Metamaterial Edge
  Coupler}.
\newblock \bibinfo{journal}{\emph{ACS Photonics}} \bibinfo{volume}{8},
  \bibinfo{number}{11} (\bibinfo{year}{2021}), \bibinfo{pages}{3226--3233}.
\newblock


\bibitem[He et~al\mbox{.}(2016)]%
        {he2016deep}
\bibfield{author}{\bibinfo{person}{Kaiming He}, \bibinfo{person}{Xiangyu
  Zhang}, \bibinfo{person}{Shaoqing Ren}, {and} \bibinfo{person}{Jian Sun}.}
  \bibinfo{year}{2016}\natexlab{}.
\newblock \showarticletitle{Deep residual learning for image recognition}. In
  \bibinfo{booktitle}{\emph{Proceedings of the IEEE conference on computer
  vision and pattern recognition}}. \bibinfo{pages}{770--778}.
\newblock


\bibitem[Hendry et~al\mbox{.}(2014)]%
        {hendry2014}
\bibfield{author}{\bibinfo{person}{Robert Hendry}, \bibinfo{person}{Dessislava
  Nikolova}, \bibinfo{person}{Sebastien Rumley}, \bibinfo{person}{Noam Ophir},
  {and} \bibinfo{person}{Keren Bergman}.} \bibinfo{year}{2014}\natexlab{}.
\newblock \showarticletitle{Physical layer analysis and modeling of silicon
  photonic WDM bus architectures}. In \bibinfo{booktitle}{\emph{Proc. HiPEAC
  Workshop}}. \bibinfo{pages}{20--22}.
\newblock


\bibitem[Hsu et~al\mbox{.}(2022)]%
        {hsu2022}
\bibfield{author}{\bibinfo{person}{Chung-Yu Hsu}, \bibinfo{person}{Gow-Zin
  Yiu}, {and} \bibinfo{person}{You-Chia Chang}.}
  \bibinfo{year}{2022}\natexlab{}.
\newblock \showarticletitle{Free-space applications of silicon photonics: A
  review}.
\newblock \bibinfo{journal}{\emph{Micromachines}} \bibinfo{volume}{13},
  \bibinfo{number}{7} (\bibinfo{year}{2022}), \bibinfo{pages}{990}.
\newblock


\bibitem[Hu(2016)]%
        {hu2016system}
\bibfield{author}{\bibinfo{person}{John Hu}.} \bibinfo{year}{2016}\natexlab{}.
\newblock \showarticletitle{System level co-cptimizations of 2.5 D/3D hybrid
  integration for high performance computing system}. In
  \bibinfo{booktitle}{\emph{Semicon West}}, Vol.~\bibinfo{volume}{2016}.
\newblock


\bibitem[Hu et~al\mbox{.}(2022)]%
        {hu}
\bibfield{author}{\bibinfo{person}{Yuhang Hu}, \bibinfo{person}{Zihao Yang},
  \bibinfo{person}{Nuo Chen}, \bibinfo{person}{Hanwen Hu},
  \bibinfo{person}{Bowen Zhang}, \bibinfo{person}{Haofan Yang},
  \bibinfo{person}{Xinda Lu}, \bibinfo{person}{Xinliang Zhang}, {and}
  \bibinfo{person}{Jing Xu}.} \bibinfo{year}{2022}\natexlab{}.
\newblock \showarticletitle{3$\times$ 40 Gbit/s All-Optical Logic Operation
  Based on Low-Loss Triple-Mode Silicon Waveguide}.
\newblock \bibinfo{journal}{\emph{Micromachines}} \bibinfo{volume}{13},
  \bibinfo{number}{1} (\bibinfo{year}{2022}), \bibinfo{pages}{90}.
\newblock


\bibitem[Iyer(2016)]%
        {iyer2016}
\bibfield{author}{\bibinfo{person}{Subramanian~S Iyer}.}
  \bibinfo{year}{2016}\natexlab{}.
\newblock \showarticletitle{Heterogeneous integration for performance and
  scaling}.
\newblock \bibinfo{journal}{\emph{IEEE Transactions on Components, Packaging
  and Manufacturing Technology}} \bibinfo{volume}{6}, \bibinfo{number}{7}
  (\bibinfo{year}{2016}), \bibinfo{pages}{973--982}.
\newblock


\bibitem[Jangam et~al\mbox{.}(2017)]%
        {jangam2017}
\bibfield{author}{\bibinfo{person}{SivaChandra Jangam},
  \bibinfo{person}{Saptadeep Pal}, \bibinfo{person}{Adeel Bajwa},
  \bibinfo{person}{Sudhakar Pamarti}, \bibinfo{person}{Puneet Gupta}, {and}
  \bibinfo{person}{Subramanian~S Iyer}.} \bibinfo{year}{2017}\natexlab{}.
\newblock \showarticletitle{Latency, bandwidth and power benefits of the
  superchips integration scheme}. In \bibinfo{booktitle}{\emph{2017 IEEE 67th
  Electronic Components and Technology Conference (ECTC)}}. IEEE,
  \bibinfo{pages}{86--94}.
\newblock


\bibitem[Jerger et~al\mbox{.}(2014)]%
        {jerger2014noc}
\bibfield{author}{\bibinfo{person}{Natalie~Enright Jerger},
  \bibinfo{person}{Ajaykumar Kannan}, \bibinfo{person}{Zimo Li}, {and}
  \bibinfo{person}{Gabriel~H Loh}.} \bibinfo{year}{2014}\natexlab{}.
\newblock \showarticletitle{Noc architectures for silicon interposer systems:
  Why pay for more wires when you can get them (from your interposer) for
  free?}. In \bibinfo{booktitle}{\emph{2014 47th Annual IEEE/ACM International
  Symposium on Microarchitecture}}. IEEE, \bibinfo{pages}{458--470}.
\newblock


\bibitem[Jouppi et~al\mbox{.}(2017)]%
        {jouppi2017}
\bibfield{author}{\bibinfo{person}{Norman~P Jouppi}, \bibinfo{person}{Cliff
  Young}, \bibinfo{person}{Nishant Patil}, \bibinfo{person}{David Patterson},
  \bibinfo{person}{Gaurav Agrawal}, \bibinfo{person}{Raminder Bajwa},
  \bibinfo{person}{Sarah Bates}, \bibinfo{person}{Suresh Bhatia},
  \bibinfo{person}{Nan Boden}, \bibinfo{person}{Al Borchers}, {et~al\mbox{.}}}
  \bibinfo{year}{2017}\natexlab{}.
\newblock \showarticletitle{In-datacenter performance analysis of a tensor
  processing unit}. In \bibinfo{booktitle}{\emph{Proceedings of the 44th annual
  international symposium on computer architecture}}. \bibinfo{pages}{1--12}.
\newblock


\bibitem[Kannan et~al\mbox{.}(2015)]%
        {kannan2015}
\bibfield{author}{\bibinfo{person}{Ajaykumar Kannan},
  \bibinfo{person}{Natalie~Enright Jerger}, {and} \bibinfo{person}{Gabriel~H
  Loh}.} \bibinfo{year}{2015}\natexlab{}.
\newblock \showarticletitle{Enabling interposer-based disintegration of
  multi-core processors}. In \bibinfo{booktitle}{\emph{2015 48th Annual
  IEEE/ACM International Symposium on Microarchitecture (MICRO)}}. IEEE,
  \bibinfo{pages}{546--558}.
\newblock


\bibitem[Karempudi et~al\mbox{.}(2020)]%
        {karempudi2020}
\bibfield{author}{\bibinfo{person}{V~Sai~Praneeth Karempudi},
  \bibinfo{person}{Sairam Sri~Vatsavayi}, {and} \bibinfo{person}{Ishan
  Thakkar}.} \bibinfo{year}{2020}\natexlab{}.
\newblock \showarticletitle{Redesigning Photonic Interconnects with
  Silicon-on-Sapphire Device Platform for Ultra-Low-Energy On-Chip
  Communication}. In \bibinfo{booktitle}{\emph{Proceedings of the 2020 on Great
  Lakes Symposium on VLSI}}. \bibinfo{pages}{247--252}.
\newblock


\bibitem[Khani et~al\mbox{.}(2021)]%
        {khani2021}
\bibfield{author}{\bibinfo{person}{Mehrdad Khani}, \bibinfo{person}{Manya
  Ghobadi}, \bibinfo{person}{Mohammad Alizadeh}, \bibinfo{person}{Ziyi Zhu},
  \bibinfo{person}{Madeleine Glick}, \bibinfo{person}{Keren Bergman},
  \bibinfo{person}{Amin Vahdat}, \bibinfo{person}{Benjamin Klenk}, {and}
  \bibinfo{person}{Eiman Ebrahimi}.} \bibinfo{year}{2021}\natexlab{}.
\newblock \showarticletitle{SiP-ML: high-bandwidth optical network
  interconnects for machine learning training}. In
  \bibinfo{booktitle}{\emph{Proceedings of the 2021 ACM SIGCOMM 2021
  Conference}}. \bibinfo{pages}{657--675}.
\newblock


\bibitem[Kim et~al\mbox{.}(2019)]%
        {kim2019}
\bibfield{author}{\bibinfo{person}{Bok~Young Kim}, \bibinfo{person}{Yoshitomo
  Okawachi}, \bibinfo{person}{Jae~K Jang}, \bibinfo{person}{Mengjie Yu},
  \bibinfo{person}{Xingchen Ji}, \bibinfo{person}{Yun Zhao},
  \bibinfo{person}{Chaitanya Joshi}, \bibinfo{person}{Michal Lipson}, {and}
  \bibinfo{person}{Alexander~L Gaeta}.} \bibinfo{year}{2019}\natexlab{}.
\newblock \showarticletitle{Turn-key, high-efficiency Kerr comb source}.
\newblock \bibinfo{journal}{\emph{Optics letters}} \bibinfo{volume}{44},
  \bibinfo{number}{18} (\bibinfo{year}{2019}), \bibinfo{pages}{4475--4478}.
\newblock


\bibitem[Lai et~al\mbox{.}(2016)]%
        {lai2016}
\bibfield{author}{\bibinfo{person}{Chieh-Lung Lai}, \bibinfo{person}{Hung-Yuan
  Li}, \bibinfo{person}{Allen Chen}, {and} \bibinfo{person}{Terren Lu}.}
  \bibinfo{year}{2016}\natexlab{}.
\newblock \showarticletitle{Silicon interposer warpage study for 2.5 D IC
  without TSV utilizing glass carrier CTE and passivation thickness tuning}. In
  \bibinfo{booktitle}{\emph{2016 IEEE 66th Electronic Components and Technology
  Conference (ECTC)}}. IEEE, \bibinfo{pages}{310--315}.
\newblock


\bibitem[Lee et~al\mbox{.}(2008)]%
        {lee2008}
\bibfield{author}{\bibinfo{person}{Benjamin~G Lee}, \bibinfo{person}{Xiaogang
  Chen}, \bibinfo{person}{Aleksandr Biberman}, \bibinfo{person}{Xiaoping Liu},
  \bibinfo{person}{I-Wei Hsieh}, \bibinfo{person}{Cheng-Yun Chou},
  \bibinfo{person}{Jerry~I Dadap}, \bibinfo{person}{Fengnian Xia},
  \bibinfo{person}{William~MJ Green}, \bibinfo{person}{Lidija Sekaric},
  {et~al\mbox{.}}} \bibinfo{year}{2008}\natexlab{}.
\newblock \showarticletitle{Ultrahigh-bandwidth silicon photonic nanowire
  waveguides for on-chip networks}.
\newblock \bibinfo{journal}{\emph{IEEE Photonics Technology Letters}}
  \bibinfo{volume}{20}, \bibinfo{number}{6} (\bibinfo{year}{2008}),
  \bibinfo{pages}{398--400}.
\newblock


\bibitem[Li and Bogaerts(2016)]%
        {li2016}
\bibfield{author}{\bibinfo{person}{Ang Li} {and} \bibinfo{person}{Wim
  Bogaerts}.} \bibinfo{year}{2016}\natexlab{}.
\newblock \showarticletitle{A simple and novel method to obtain an FSR free
  silicon ring resonator}. In \bibinfo{booktitle}{\emph{Silicon Photonics and
  Photonic Integrated Circuits V}}, Vol.~\bibinfo{volume}{9891}. International
  Society for Optics and Photonics, \bibinfo{pages}{989115}.
\newblock


\bibitem[Li et~al\mbox{.}(2012)]%
        {Qi2012}
\bibfield{author}{\bibinfo{person}{Qi Li}, \bibinfo{person}{Noam Ophir},
  \bibinfo{person}{Lin Xu}, \bibinfo{person}{Kishore Padmaraju},
  \bibinfo{person}{Long Chen}, \bibinfo{person}{Michal Lipson}, {and}
  \bibinfo{person}{Keren Bergman}.} \bibinfo{year}{2012}\natexlab{}.
\newblock \showarticletitle{Experimental characterization of the optical-power
  upper bound in a silicon microring modulator}. In
  \bibinfo{booktitle}{\emph{2012 Optical Interconnects Conference}}. IEEE,
  \bibinfo{pages}{38--39}.
\newblock


\bibitem[Luo et~al\mbox{.}(2012)]%
        {luo2012}
\bibfield{author}{\bibinfo{person}{Lian-Wee Luo}, \bibinfo{person}{Gustavo~S
  Wiederhecker}, \bibinfo{person}{Kyle Preston}, {and} \bibinfo{person}{Michal
  Lipson}.} \bibinfo{year}{2012}\natexlab{}.
\newblock \showarticletitle{Power insensitive silicon microring resonators}.
\newblock \bibinfo{journal}{\emph{Optics letters}} \bibinfo{volume}{37},
  \bibinfo{number}{4} (\bibinfo{year}{2012}), \bibinfo{pages}{590--592}.
\newblock


\bibitem[Mahajan et~al\mbox{.}(2016)]%
        {mahajan2016}
\bibfield{author}{\bibinfo{person}{Ravi Mahajan}, \bibinfo{person}{Robert
  Sankman}, \bibinfo{person}{Neha Patel}, \bibinfo{person}{Dae-Woo Kim},
  \bibinfo{person}{Kemal Aygun}, \bibinfo{person}{Zhiguo Qian},
  \bibinfo{person}{Yidnekachew Mekonnen}, \bibinfo{person}{Islam Salama},
  \bibinfo{person}{Sujit Sharan}, \bibinfo{person}{Deepti Iyengar},
  {et~al\mbox{.}}} \bibinfo{year}{2016}\natexlab{}.
\newblock \showarticletitle{Embedded multi-die interconnect bridge (EMIB)--a
  high density, high bandwidth packaging interconnect}. In
  \bibinfo{booktitle}{\emph{2016 IEEE 66th Electronic Components and Technology
  Conference (ECTC)}}. IEEE, \bibinfo{pages}{557--565}.
\newblock


\bibitem[Mistry et~al\mbox{.}(2020)]%
        {mistry2020}
\bibfield{author}{\bibinfo{person}{Ajay Mistry}, \bibinfo{person}{Mustafa
  Hammood}, \bibinfo{person}{Hossam Shoman}, \bibinfo{person}{Stephen Lin},
  \bibinfo{person}{Lukas Chrostowski}, {and} \bibinfo{person}{Nicolas~AF
  Jaeger}.} \bibinfo{year}{2020}\natexlab{}.
\newblock \showarticletitle{Free-spectral-range-free microring-based coupling
  modulator with integrated contra-directional-couplers}. In
  \bibinfo{booktitle}{\emph{Optical Components and Materials XVII}},
  Vol.~\bibinfo{volume}{11276}. International Society for Optics and Photonics,
  \bibinfo{pages}{1127607}.
\newblock


\bibitem[Morichetti et~al\mbox{.}(2021)]%
        {morichetti2021}
\bibfield{author}{\bibinfo{person}{Francesco Morichetti},
  \bibinfo{person}{Maziyar Milanizadeh}, \bibinfo{person}{Matteo Petrini},
  \bibinfo{person}{Francesco Zanetto}, \bibinfo{person}{Giorgio Ferrari},
  \bibinfo{person}{Douglas~Oliveira de Aguiar}, \bibinfo{person}{Emanuele
  Guglielmi}, \bibinfo{person}{Marco Sampietro}, {and} \bibinfo{person}{Andrea
  Melloni}.} \bibinfo{year}{2021}\natexlab{}.
\newblock \showarticletitle{Polarization-transparent silicon photonic add-drop
  multiplexer with wideband hitless tuneability}.
\newblock \bibinfo{journal}{\emph{Nature Communications}} \bibinfo{volume}{12},
  \bibinfo{number}{1} (\bibinfo{year}{2021}), \bibinfo{pages}{1--7}.
\newblock


\bibitem[Mu et~al\mbox{.}(2020)]%
        {mu2020}
\bibfield{author}{\bibinfo{person}{Xin Mu}, \bibinfo{person}{Sailong Wu},
  \bibinfo{person}{Lirong Cheng}, {and} \bibinfo{person}{HY Fu}.}
  \bibinfo{year}{2020}\natexlab{}.
\newblock \showarticletitle{Edge couplers in silicon photonic integrated
  circuits: A review}.
\newblock \bibinfo{journal}{\emph{Applied Sciences}} \bibinfo{volume}{10},
  \bibinfo{number}{4} (\bibinfo{year}{2020}), \bibinfo{pages}{1538}.
\newblock


\bibitem[Naffziger et~al\mbox{.}(2021)]%
        {naffziger2021pioneering}
\bibfield{author}{\bibinfo{person}{Samuel Naffziger}, \bibinfo{person}{Noah
  Beck}, \bibinfo{person}{Thomas Burd}, \bibinfo{person}{Kevin Lepak},
  \bibinfo{person}{Gabriel~H Loh}, \bibinfo{person}{Mahesh Subramony}, {and}
  \bibinfo{person}{Sean White}.} \bibinfo{year}{2021}\natexlab{}.
\newblock \showarticletitle{Pioneering Chiplet Technology and Design for the
  AMD EPYC™ and Ryzen™ Processor Families: Industrial Product}. In
  \bibinfo{booktitle}{\emph{2021 ACM/IEEE 48th Annual International Symposium
  on Computer Architecture (ISCA)}}. IEEE, \bibinfo{pages}{57--70}.
\newblock


\bibitem[OpenAI(2018)]%
        {AIandcompute}
\bibfield{author}{\bibinfo{person}{OpenAI}.} \bibinfo{year}{2018}\natexlab{}.
\newblock \bibinfo{booktitle}{}.
\newblock


\bibitem[Ophir et~al\mbox{.}(2010)]%
        {ophir2010}
\bibfield{author}{\bibinfo{person}{Noam Ophir}, \bibinfo{person}{Aleksandr
  Biberman}, \bibinfo{person}{Jacob~S Levy}, \bibinfo{person}{Kishore
  Padmaraju}, \bibinfo{person}{Kevin~J Luke}, \bibinfo{person}{Michal Lipson},
  {and} \bibinfo{person}{Keren Bergman}.} \bibinfo{year}{2010}\natexlab{}.
\newblock \showarticletitle{Demonstration of 1.28-Tb/s transmission in
  next-generation nanowires for photonic networks-on-chip}. In
  \bibinfo{booktitle}{\emph{2010 23rd Annual Meeting of the IEEE Photonics
  Society}}. IEEE, \bibinfo{pages}{560--561}.
\newblock


\bibitem[Pal et~al\mbox{.}(2021)]%
        {pal2021}
\bibfield{author}{\bibinfo{person}{Saptadeep Pal}, \bibinfo{person}{Jingyang
  Liu}, \bibinfo{person}{Irina Alam}, \bibinfo{person}{Nicholas Cebry},
  \bibinfo{person}{Haris Suhail}, \bibinfo{person}{Shi Bu},
  \bibinfo{person}{Subramanian~S Iyer}, \bibinfo{person}{Sudhakar Pamarti},
  \bibinfo{person}{Rakesh Kumar}, {and} \bibinfo{person}{Puneet Gupta}.}
  \bibinfo{year}{2021}\natexlab{}.
\newblock \showarticletitle{Designing a 2048-Chiplet, 14336-Core Waferscale
  Processor}. In \bibinfo{booktitle}{\emph{2021 58th ACM/IEEE Design Automation
  Conference (DAC)}}. IEEE, \bibinfo{pages}{1183--1188}.
\newblock


\bibitem[Pal et~al\mbox{.}(2018)]%
        {pal2018}
\bibfield{author}{\bibinfo{person}{Saptadeep Pal}, \bibinfo{person}{Daniel
  Petrisko}, \bibinfo{person}{Adeel~A Bajwa}, \bibinfo{person}{Puneet Gupta},
  \bibinfo{person}{Subramanian~S Iyer}, {and} \bibinfo{person}{Rakesh Kumar}.}
  \bibinfo{year}{2018}\natexlab{}.
\newblock \showarticletitle{A case for packageless processors}. In
  \bibinfo{booktitle}{\emph{2018 IEEE international symposium on high
  performance computer architecture (HPCA)}}. IEEE, \bibinfo{pages}{466--479}.
\newblock


\bibitem[Pal et~al\mbox{.}(2019)]%
        {pal2019}
\bibfield{author}{\bibinfo{person}{Saptadeep Pal}, \bibinfo{person}{Daniel
  Petrisko}, \bibinfo{person}{Matthew Tomei}, \bibinfo{person}{Puneet Gupta},
  \bibinfo{person}{Subramanian~S Iyer}, {and} \bibinfo{person}{Rakesh Kumar}.}
  \bibinfo{year}{2019}\natexlab{}.
\newblock \showarticletitle{Architecting waferscale processors-a GPU case
  study}. In \bibinfo{booktitle}{\emph{2019 IEEE International Symposium on
  High Performance Computer Architecture (HPCA)}}. IEEE,
  \bibinfo{pages}{250--263}.
\newblock


\bibitem[Pan et~al\mbox{.}(2010)]%
        {pan2010}
\bibfield{author}{\bibinfo{person}{Yan Pan}, \bibinfo{person}{John Kim}, {and}
  \bibinfo{person}{Gokhan Memik}.} \bibinfo{year}{2010}\natexlab{}.
\newblock \showarticletitle{Flexishare: Channel sharing for an energy-efficient
  nanophotonic crossbar}. In \bibinfo{booktitle}{\emph{HPCA-16 2010 The
  Sixteenth International Symposium on High-Performance Computer
  Architecture}}. IEEE, \bibinfo{pages}{1--12}.
\newblock


\bibitem[Pan et~al\mbox{.}(2009)]%
        {pan2009}
\bibfield{author}{\bibinfo{person}{Yan Pan}, \bibinfo{person}{Prabhat Kumar},
  \bibinfo{person}{John Kim}, \bibinfo{person}{Gokhan Memik},
  \bibinfo{person}{Yu Zhang}, {and} \bibinfo{person}{Alok Choudhary}.}
  \bibinfo{year}{2009}\natexlab{}.
\newblock \showarticletitle{Firefly: Illuminating future network-on-chip with
  nanophotonics}. In \bibinfo{booktitle}{\emph{Proceedings of the 36th annual
  international symposium on Computer architecture}}.
  \bibinfo{pages}{429--440}.
\newblock


\bibitem[Pasricha and Nikdast(2020)]%
        {pasricha2020}
\bibfield{author}{\bibinfo{person}{Sudeep Pasricha} {and}
  \bibinfo{person}{Mahdi Nikdast}.} \bibinfo{year}{2020}\natexlab{}.
\newblock \showarticletitle{A survey of silicon photonics for energy-efficient
  manycore computing}.
\newblock \bibinfo{journal}{\emph{IEEE Design \& Test}} \bibinfo{volume}{37},
  \bibinfo{number}{4} (\bibinfo{year}{2020}), \bibinfo{pages}{60--81}.
\newblock


\bibitem[Petrini et~al\mbox{.}(2021)]%
        {petrini2021}
\bibfield{author}{\bibinfo{person}{Matteo Petrini}, \bibinfo{person}{Maziyar
  Milanizadeh}, \bibinfo{person}{Francesco Zanetto}, \bibinfo{person}{Giorgio
  Ferrari}, \bibinfo{person}{Marco Sampietro}, \bibinfo{person}{Francesco
  Morichetti}, {and} \bibinfo{person}{Andrea Melloni}.}
  \bibinfo{year}{2021}\natexlab{}.
\newblock \showarticletitle{Reconfigurable FSR-free microring resonator filter
  with wide hitless tunability}. In \bibinfo{booktitle}{\emph{2021 IEEE
  Photonics Society Summer Topicals Meeting Series (SUM)}}. IEEE,
  \bibinfo{pages}{1--2}.
\newblock


\bibitem[Rahim et~al\mbox{.}(2017)]%
        {rahim2017}
\bibfield{author}{\bibinfo{person}{Abdul Rahim}, \bibinfo{person}{Eva
  Ryckeboer}, \bibinfo{person}{Ananth~Z Subramanian},
  \bibinfo{person}{St{\'e}phane Clemmen}, \bibinfo{person}{Bart Kuyken},
  \bibinfo{person}{Ashim Dhakal}, \bibinfo{person}{Ali Raza},
  \bibinfo{person}{Artur Hermans}, \bibinfo{person}{Muhammad Muneeb},
  \bibinfo{person}{S{\"o}ren Dhoore}, {et~al\mbox{.}}}
  \bibinfo{year}{2017}\natexlab{}.
\newblock \showarticletitle{Expanding the silicon photonics portfolio with
  silicon nitride photonic integrated circuits}.
\newblock \bibinfo{journal}{\emph{Journal of lightwave technology}}
  \bibinfo{volume}{35}, \bibinfo{number}{4} (\bibinfo{year}{2017}),
  \bibinfo{pages}{639--649}.
\newblock


\bibitem[Rakowski et~al\mbox{.}(2018)]%
        {rakowski2018}
\bibfield{author}{\bibinfo{person}{Michal Rakowski}, \bibinfo{person}{Yoojin
  Ban}, \bibinfo{person}{Peter De~Heyn}, \bibinfo{person}{Nicolas Pantano},
  \bibinfo{person}{Brad Snyder}, \bibinfo{person}{Sadhishkumar Balakrishnan},
  \bibinfo{person}{Stefaan Van~Huylenbroeck}, \bibinfo{person}{Lieve Bogaerts},
  \bibinfo{person}{Caroline Demeurisse}, \bibinfo{person}{Fumihiro Inoue},
  {et~al\mbox{.}}} \bibinfo{year}{2018}\natexlab{}.
\newblock \showarticletitle{Hybrid 14nm FinFET-Silicon Photonics Technology for
  Low-Power Tb/s/mm 2 Optical I/O}. In \bibinfo{booktitle}{\emph{2018 IEEE
  Symposium on VLSI Technology}}. IEEE, \bibinfo{pages}{221--222}.
\newblock


\bibitem[Rizzo et~al\mbox{.}(2019)]%
        {rizzo2019}
\bibfield{author}{\bibinfo{person}{Anthony Rizzo}, \bibinfo{person}{Yanir
  London}, \bibinfo{person}{Geza Kurczveil}, \bibinfo{person}{Thomas
  Van~Vaerenbergh}, \bibinfo{person}{Marco Fiorentino}, \bibinfo{person}{Ashkan
  Seyedi}, \bibinfo{person}{Daniil Livshits}, \bibinfo{person}{Raymond~G
  Beausoleil}, {and} \bibinfo{person}{Keren Bergman}.}
  \bibinfo{year}{2019}\natexlab{}.
\newblock \showarticletitle{Energy efficiency analysis of frequency comb
  sources for silicon photonic interconnects}. In
  \bibinfo{booktitle}{\emph{2019 IEEE Optical Interconnects Conference (OI)}}.
  IEEE, \bibinfo{pages}{1--2}.
\newblock


\bibitem[Sai Praneeth~Karempudi et~al\mbox{.}(2021)]%
        {sai2021}
\bibfield{author}{\bibinfo{person}{Venkata Sai Praneeth~Karempudi},
  \bibinfo{person}{Febin Sunny}, \bibinfo{person}{Ishan~G Thakkar},
  \bibinfo{person}{Sai Vineel Reddy~Chittamuru}, \bibinfo{person}{Mahdi
  Nikdast}, {and} \bibinfo{person}{Sudeep Pasricha}.}
  \bibinfo{year}{2021}\natexlab{}.
\newblock \showarticletitle{Photonic Networks-on-Chip Employing Multilevel
  Signaling: A Cross-Layer Comparative Study}.
\newblock \bibinfo{journal}{\emph{arXiv e-prints}} (\bibinfo{year}{2021}),
  \bibinfo{pages}{arXiv--2110}.
\newblock


\bibitem[S{\'a}nchez-Postigo et~al\mbox{.}(2021)]%
        {sanchez2021}
\bibfield{author}{\bibinfo{person}{Alejandro S{\'a}nchez-Postigo},
  \bibinfo{person}{Robert Halir}, \bibinfo{person}{J~Gonzalo
  Wang{\"u}emert-P{\'e}rez}, \bibinfo{person}{Alejandro Ortega-Mo{\~n}ux},
  \bibinfo{person}{Shurui Wang}, \bibinfo{person}{Martin Vachon},
  \bibinfo{person}{Jens~H Schmid}, \bibinfo{person}{Dan-Xia Xu},
  \bibinfo{person}{Pavel Cheben}, {and} \bibinfo{person}{{\'I}{\~n}igo
  Molina-Fern{\'a}ndez}.} \bibinfo{year}{2021}\natexlab{}.
\newblock \showarticletitle{Breaking the coupling efficiency--bandwidth
  trade-off in surface grating couplers using zero-order radiation}.
\newblock \bibinfo{journal}{\emph{Laser \& Photonics Reviews}}
  \bibinfo{volume}{15}, \bibinfo{number}{6} (\bibinfo{year}{2021}),
  \bibinfo{pages}{2000542}.
\newblock


\bibitem[Sarangi et~al\mbox{.}(2015)]%
        {sarangi2015}
\bibfield{author}{\bibinfo{person}{Smruti~R Sarangi},
  \bibinfo{person}{Rajshekar Kalayappan}, \bibinfo{person}{Prathmesh
  Kallurkar}, \bibinfo{person}{Seep Goel}, {and} \bibinfo{person}{Eldhose
  Peter}.} \bibinfo{year}{2015}\natexlab{}.
\newblock \showarticletitle{Tejas: A java based versatile micro-architectural
  simulator}. In \bibinfo{booktitle}{\emph{2015 25th International Workshop on
  Power and Timing Modeling, Optimization and Simulation (PATMOS)}}. IEEE,
  \bibinfo{pages}{47--54}.
\newblock


\bibitem[Shallue et~al\mbox{.}(2018)]%
        {shallue2018measuring}
\bibfield{author}{\bibinfo{person}{Christopher~J Shallue},
  \bibinfo{person}{Jaehoon Lee}, \bibinfo{person}{Joseph Antognini},
  \bibinfo{person}{Jascha Sohl-Dickstein}, \bibinfo{person}{Roy Frostig}, {and}
  \bibinfo{person}{George~E Dahl}.} \bibinfo{year}{2018}\natexlab{}.
\newblock \showarticletitle{Measuring the effects of data parallelism on neural
  network training}.
\newblock \bibinfo{journal}{\emph{arXiv preprint arXiv:1811.03600}}
  (\bibinfo{year}{2018}).
\newblock


\bibitem[Shoeybi et~al\mbox{.}(2019)]%
        {shoeybi2019megatron}
\bibfield{author}{\bibinfo{person}{Mohammad Shoeybi}, \bibinfo{person}{Mostofa
  Patwary}, \bibinfo{person}{Raul Puri}, \bibinfo{person}{Patrick LeGresley},
  \bibinfo{person}{Jared Casper}, {and} \bibinfo{person}{Bryan Catanzaro}.}
  \bibinfo{year}{2019}\natexlab{}.
\newblock \showarticletitle{Megatron-lm: Training multi-billion parameter
  language models using model parallelism}.
\newblock \bibinfo{journal}{\emph{arXiv preprint arXiv:1909.08053}}
  (\bibinfo{year}{2019}).
\newblock


\bibitem[Sodani et~al\mbox{.}(2016)]%
        {sodani2016}
\bibfield{author}{\bibinfo{person}{Avinash Sodani}, \bibinfo{person}{Roger
  Gramunt}, \bibinfo{person}{Jesus Corbal}, \bibinfo{person}{Ho-Seop Kim},
  \bibinfo{person}{Krishna Vinod}, \bibinfo{person}{Sundaram Chinthamani},
  \bibinfo{person}{Steven Hutsell}, \bibinfo{person}{Rajat Agarwal}, {and}
  \bibinfo{person}{Yen-Chen Liu}.} \bibinfo{year}{2016}\natexlab{}.
\newblock \showarticletitle{Knights landing: Second-generation intel xeon phi
  product}.
\newblock \bibinfo{journal}{\emph{Ieee micro}} \bibinfo{volume}{36},
  \bibinfo{number}{2} (\bibinfo{year}{2016}), \bibinfo{pages}{34--46}.
\newblock


\bibitem[Stern et~al\mbox{.}(2018)]%
        {stern2018}
\bibfield{author}{\bibinfo{person}{Brian Stern}, \bibinfo{person}{Xingchen Ji},
  \bibinfo{person}{Yoshitomo Okawachi}, \bibinfo{person}{Alexander~L Gaeta},
  {and} \bibinfo{person}{Michal Lipson}.} \bibinfo{year}{2018}\natexlab{}.
\newblock \showarticletitle{Battery-operated integrated frequency comb
  generator}.
\newblock \bibinfo{journal}{\emph{Nature}} \bibinfo{volume}{562},
  \bibinfo{number}{7727} (\bibinfo{year}{2018}), \bibinfo{pages}{401--405}.
\newblock


\bibitem[Stojanovi{\'c} et~al\mbox{.}(2018)]%
        {stojanovic2018}
\bibfield{author}{\bibinfo{person}{Vladimir Stojanovi{\'c}},
  \bibinfo{person}{Rajeev~J Ram}, \bibinfo{person}{Milos Popovi{\'c}},
  \bibinfo{person}{Sen Lin}, \bibinfo{person}{Sajjad Moazeni},
  \bibinfo{person}{Mark Wade}, \bibinfo{person}{Chen Sun},
  \bibinfo{person}{Luca Alloatti}, \bibinfo{person}{Amir Atabaki},
  \bibinfo{person}{Fabio Pavanello}, {et~al\mbox{.}}}
  \bibinfo{year}{2018}\natexlab{}.
\newblock \showarticletitle{Monolithic silicon-photonic platforms in
  state-of-the-art CMOS SOI processes}.
\newblock \bibinfo{journal}{\emph{Optics express}} \bibinfo{volume}{26},
  \bibinfo{number}{10} (\bibinfo{year}{2018}), \bibinfo{pages}{13106--13121}.
\newblock


\bibitem[Stow et~al\mbox{.}(2017)]%
        {stow2017}
\bibfield{author}{\bibinfo{person}{Dylan Stow}, \bibinfo{person}{Yuan Xie},
  \bibinfo{person}{Taniya Siddiqua}, {and} \bibinfo{person}{Gabriel~H Loh}.}
  \bibinfo{year}{2017}\natexlab{}.
\newblock \showarticletitle{Cost-effective design of scalable high-performance
  systems using active and passive interposers}. In
  \bibinfo{booktitle}{\emph{2017 IEEE/ACM International Conference on
  Computer-Aided Design (ICCAD)}}. IEEE, \bibinfo{pages}{728--735}.
\newblock


\bibitem[Sun et~al\mbox{.}(2019)]%
        {sun2019}
\bibfield{author}{\bibinfo{person}{Peng Sun}, \bibinfo{person}{Jared Hulme},
  \bibinfo{person}{Thomas Van~Vaerenbergh}, \bibinfo{person}{Jinsoo Rhim},
  \bibinfo{person}{Charles Baudot}, \bibinfo{person}{Frederic Boeuf},
  \bibinfo{person}{Nathalie Vulliet}, \bibinfo{person}{Ashkan Seyedi},
  \bibinfo{person}{Marco Fiorentino}, {and} \bibinfo{person}{Raymond~G
  Beausoleil}.} \bibinfo{year}{2019}\natexlab{}.
\newblock \showarticletitle{Statistical behavioral models of silicon ring
  resonators at a commercial CMOS foundry}.
\newblock \bibinfo{journal}{\emph{IEEE Journal of Selected Topics in Quantum
  Electronics}} \bibinfo{volume}{26}, \bibinfo{number}{2}
  (\bibinfo{year}{2019}), \bibinfo{pages}{1--10}.
\newblock


\bibitem[Thakkar et~al\mbox{.}(2016)]%
        {thakkar2016}
\bibfield{author}{\bibinfo{person}{Ishan~G Thakkar}, \bibinfo{person}{Sai
  Vineel~Reddy Chittamuru}, {and} \bibinfo{person}{Sudeep Pasricha}.}
  \bibinfo{year}{2016}\natexlab{}.
\newblock \showarticletitle{A comparative analysis of front-end and back-end
  compatible silicon photonic on-chip interconnects}. In
  \bibinfo{booktitle}{\emph{2016 ACM/IEEE International Workshop on System
  Level Interconnect Prediction (SLIP)}}. IEEE, \bibinfo{pages}{1--8}.
\newblock


\bibitem[Thakkar et~al\mbox{.}(2017)]%
        {thakkar2017}
\bibfield{author}{\bibinfo{person}{Ishan~G Thakkar}, \bibinfo{person}{Sai
  Vineel~Reddy Chittamuru}, {and} \bibinfo{person}{Sudeep Pasricha}.}
  \bibinfo{year}{2017}\natexlab{}.
\newblock \showarticletitle{Improving the reliability and energy-efficiency of
  high-bandwidth photonic NoC architectures with multilevel signaling}. In
  \bibinfo{booktitle}{\emph{2017 Eleventh IEEE/ACM International Symposium on
  Networks-on-Chip (NOCS)}}. IEEE, \bibinfo{pages}{1--8}.
\newblock


\bibitem[Thonnart et~al\mbox{.}(2020)]%
        {thonnart2020}
\bibfield{author}{\bibinfo{person}{Yvain Thonnart},
  \bibinfo{person}{St{\'e}phane Bernab{\'e}}, \bibinfo{person}{Jean
  Charbonnier}, \bibinfo{person}{Christian Bernard}, \bibinfo{person}{David
  Coriat}, \bibinfo{person}{C{\'e}sar Fuguet}, \bibinfo{person}{Pierre
  Tissier}, \bibinfo{person}{Beno{\^\i}t Charbonnier},
  \bibinfo{person}{St{\'e}phane Malhouitre}, \bibinfo{person}{Damien
  Saint-Patrice}, {et~al\mbox{.}}} \bibinfo{year}{2020}\natexlab{}.
\newblock \showarticletitle{POPSTAR: A robust modular optical NoC architecture
  for chiplet-based 3D integrated systems}. In \bibinfo{booktitle}{\emph{2020
  Design, Automation \& Test in Europe Conference \& Exhibition (DATE)}}. IEEE,
  \bibinfo{pages}{1456--1461}.
\newblock


\bibitem[Urbonas et~al\mbox{.}(2015)]%
        {urbonas2015}
\bibfield{author}{\bibinfo{person}{Darius Urbonas}, \bibinfo{person}{Armandas
  Bal{\v{c}}ytis}, \bibinfo{person}{Martynas Gabalis},
  \bibinfo{person}{Konstantinas Va{\v{s}}kevi{\v{c}}ius},
  \bibinfo{person}{Greta Naujokait{\.e}}, \bibinfo{person}{Saulius Juodkazis},
  {and} \bibinfo{person}{Raimondas Petru{\v{s}}kevi{\v{c}}ius}.}
  \bibinfo{year}{2015}\natexlab{}.
\newblock \showarticletitle{Ultra-wide free spectral range, enhanced
  sensitivity, and removed mode splitting SOI optical ring resonator with
  dispersive metal nanodisks}.
\newblock \bibinfo{journal}{\emph{Optics letters}} \bibinfo{volume}{40},
  \bibinfo{number}{13} (\bibinfo{year}{2015}), \bibinfo{pages}{2977--2980}.
\newblock


\bibitem[Vatsavai et~al\mbox{.}(2020)]%
        {vatsavai2020}
\bibfield{author}{\bibinfo{person}{Sairam~Sri Vatsavai},
  \bibinfo{person}{Venkata Sai~Praneeth Karempudi}, {and}
  \bibinfo{person}{Ishan Thakkar}.} \bibinfo{year}{2020}\natexlab{}.
\newblock \showarticletitle{PROTEUS: Rule-based self-adaptation in photonic
  NoCs for loss-aware co-management of laser power and performance}. In
  \bibinfo{booktitle}{\emph{2020 14th IEEE/ACM International Symposium on
  Networks-on-Chip (NOCS)}}. IEEE, \bibinfo{pages}{1--8}.
\newblock


\bibitem[Venkataramani et~al\mbox{.}(2017)]%
        {venkataramani2017}
\bibfield{author}{\bibinfo{person}{Swagath Venkataramani},
  \bibinfo{person}{Ashish Ranjan}, \bibinfo{person}{Subarno Banerjee},
  \bibinfo{person}{Dipankar Das}, \bibinfo{person}{Sasikanth Avancha},
  \bibinfo{person}{Ashok Jagannathan}, \bibinfo{person}{Ajaya Durg},
  \bibinfo{person}{Dheemanth Nagaraj}, \bibinfo{person}{Bharat Kaul},
  \bibinfo{person}{Pradeep Dubey}, {et~al\mbox{.}}}
  \bibinfo{year}{2017}\natexlab{}.
\newblock \showarticletitle{Scaledeep: A scalable compute architecture for
  learning and evaluating deep networks}. In
  \bibinfo{booktitle}{\emph{Proceedings of the 44th Annual International
  Symposium on Computer Architecture}}. \bibinfo{pages}{13--26}.
\newblock


\bibitem[Wang et~al\mbox{.}(2020)]%
        {wang2020}
\bibfield{author}{\bibinfo{person}{Yuyang Wang}, \bibinfo{person}{Jared Hulme},
  \bibinfo{person}{Peng Sun}, \bibinfo{person}{Mudit Jain},
  \bibinfo{person}{M~Ashkan Seyedi}, \bibinfo{person}{Marco Fiorentino},
  \bibinfo{person}{Raymond~G Beausoleil}, {and} \bibinfo{person}{Kwang-Ting
  Cheng}.} \bibinfo{year}{2020}\natexlab{}.
\newblock \showarticletitle{Characterization and applications of spatial
  variation models for silicon microring-based optical transceivers}. In
  \bibinfo{booktitle}{\emph{2020 57th ACM/IEEE Design Automation Conference
  (DAC)}}. IEEE, \bibinfo{pages}{1--6}.
\newblock


\bibitem[Wang et~al\mbox{.}(2018)]%
        {wang2018}
\bibfield{author}{\bibinfo{person}{Yuyang Wang}, \bibinfo{person}{M~Ashkan
  Seyedi}, \bibinfo{person}{Rui Wu}, \bibinfo{person}{Jared Hulme},
  \bibinfo{person}{Marco Fiorentino}, \bibinfo{person}{Raymond~G Beausoleil},
  {and} \bibinfo{person}{Kwang-Ting Cheng}.} \bibinfo{year}{2018}\natexlab{}.
\newblock \showarticletitle{Energy-efficient channel alignment of dwdm silicon
  photonic transceivers}. In \bibinfo{booktitle}{\emph{2018 Design, Automation
  \& Test in Europe Conference \& Exhibition (DATE)}}. IEEE,
  \bibinfo{pages}{601--604}.
\newblock


\bibitem[Wilmart et~al\mbox{.}(2019)]%
        {wilmart2019}
\bibfield{author}{\bibinfo{person}{Quentin Wilmart}, \bibinfo{person}{Houssein
  El~Dirani}, \bibinfo{person}{Nicola Tyler}, \bibinfo{person}{Daivid Fowler},
  \bibinfo{person}{St{\'e}phane Malhouitre}, \bibinfo{person}{St{\'e}phanie
  Garcia}, \bibinfo{person}{Marco Casale}, \bibinfo{person}{S{\'e}bastien
  Kerdiles}, \bibinfo{person}{Karim Hassan}, \bibinfo{person}{Christelle
  Monat}, {et~al\mbox{.}}} \bibinfo{year}{2019}\natexlab{}.
\newblock \showarticletitle{A versatile silicon-silicon nitride photonics
  platform for enhanced functionalities and applications}.
\newblock \bibinfo{journal}{\emph{Applied Sciences}} \bibinfo{volume}{9},
  \bibinfo{number}{2} (\bibinfo{year}{2019}), \bibinfo{pages}{255}.
\newblock


\bibitem[Xue et~al\mbox{.}(2017)]%
        {xue2017}
\bibfield{author}{\bibinfo{person}{Xiaoxiao Xue}, \bibinfo{person}{Pei-Hsun
  Wang}, \bibinfo{person}{Yi Xuan}, \bibinfo{person}{Minghao Qi}, {and}
  \bibinfo{person}{Andrew~M Weiner}.} \bibinfo{year}{2017}\natexlab{}.
\newblock \showarticletitle{Microresonator Kerr frequency combs with high
  conversion efficiency}.
\newblock \bibinfo{journal}{\emph{Laser \& Photonics Reviews}}
  \bibinfo{volume}{11}, \bibinfo{number}{1} (\bibinfo{year}{2017}),
  \bibinfo{pages}{1600276}.
\newblock


\bibitem[Younis et~al\mbox{.}(2018)]%
        {younis2018}
\bibfield{author}{\bibinfo{person}{Usman Younis}, \bibinfo{person}{Xianshu
  Luo}, \bibinfo{person}{Bowei Dong}, \bibinfo{person}{Li Huang},
  \bibinfo{person}{Sudheer~K Vanga}, \bibinfo{person}{Andy Eu-Jin Lim},
  \bibinfo{person}{Patrick Guo-Qiang Lo}, \bibinfo{person}{Chengkuo Lee},
  \bibinfo{person}{Andrew~A Bettiol}, {and} \bibinfo{person}{Kah-Wee Ang}.}
  \bibinfo{year}{2018}\natexlab{}.
\newblock \showarticletitle{Towards low-loss waveguides in SOI and Ge-on-SOI
  for mid-IR sensing}.
\newblock \bibinfo{journal}{\emph{Journal of Physics Communications}}
  \bibinfo{volume}{2}, \bibinfo{number}{4} (\bibinfo{year}{2018}),
  \bibinfo{pages}{045029}.
\newblock


\bibitem[Zhou et~al\mbox{.}(2017)]%
        {zhou2017}
\bibfield{author}{\bibinfo{person}{Huanying Zhou}, \bibinfo{person}{Ciyuan
  Qiu}, \bibinfo{person}{Xinhong Jiang}, \bibinfo{person}{Qingming Zhu},
  \bibinfo{person}{Yu He}, \bibinfo{person}{Yong Zhang}, \bibinfo{person}{Yikai
  Su}, {and} \bibinfo{person}{Richard Soref}.} \bibinfo{year}{2017}\natexlab{}.
\newblock \showarticletitle{Compact, submilliwatt, 2$\times$ 2 silicon
  thermo-optic switch based on photonic crystal nanobeam cavities}.
\newblock \bibinfo{journal}{\emph{Photonics Research}} \bibinfo{volume}{5},
  \bibinfo{number}{2} (\bibinfo{year}{2017}), \bibinfo{pages}{108--112}.
\newblock


\end{thebibliography}

\end{document}